\documentclass[fleqn,usenatbib]{mnras}

\usepackage{newtxtext,newtxmath}

\usepackage[T1]{fontenc}
\usepackage{ae,aecompl}

\usepackage{graphicx}	
\usepackage{amsmath}	

\def\msun{\hbox{M$_\odot$}}

\usepackage[dvipsnames]{xcolor}
\usepackage{ulem}

\title[AMR in E-MOSAICS]{Linking globular cluster formation at low and high redshift through the age-metallicity relation in E-MOSAICS}

\author[Horta et al.]{Danny Horta$^{1}$\thanks{E-mail: D.HortaDarrington@2018.ljmu.ac.uk}%
 , Meghan E. Hughes$^{1}$, Joel L. Pfeffer$^{1}$, Nate Bastian$^{1}$,
 \newauthor 
 J.~M.~Diederik Kruijssen$^{2}$, Marta Reina-Campos$^{2}$, Robert A. Crain$^{1}$
\\
$^{1}$Astrophysics Research Institute, Liverpool John Moores University, 146 Brownlow Hill, Liverpool L3 5RF, UK\\
$^2$Astronomisches Rechen-Institut, Zentrum f\"{u}r Astronomie der Universit\"{a}t Heidelberg, M\"{o}nchhofstra\ss e 12-14, 69120 Heidelberg, Germany\\
}
\date{Accepted XXX. Received YYY; in original form ZZZ}

\pubyear{2020}

\begin{document}
\label{firstpage}
\pagerange{\pageref{firstpage}--\pageref{lastpage}}
\maketitle

\begin{abstract}
We set out to compare the age-metallicity relation (AMR) of massive clusters from Magellanic Cloud mass galaxies in the E-MOSAICS suite of numerical cosmological simulations with an amalgamation of observational data of massive clusters in the  Large and Small Magellanic Clouds (LMC/SMC). We aim to test if: $i$) star cluster formation proceeds according to universal physical processes, suggestive of a common formation mechanism for young-massive clusters (YMCs), intermediate-age clusters (IACs), and ancient globular clusters (GCs); $ii$) massive clusters of all ages trace a continuous AMR; $iii$) the AMRs of smaller mass galaxies show a shallower relation when compared to more massive galaxies. Our results show that, within the uncertainties, the predicted AMRs of L/SMC-mass galaxies with similar star formation histories to the L/SMC follow the same relation as observations. We also find that the metallicity at which the AMR saturates increases with galaxy mass, which is also found for the field star AMRs. This suggests that relatively low-metallicity clusters can still form in dwarfs galaxies. Given our results, we suggest that ancient GCs share their formation mechanism with IACs and YMCs, in which GCs are the result of a universal process of star cluster formation during the early episodes of star formation in their host galaxies.

\end{abstract}



\section{Introduction}

Where and when globular clusters (GCs) form, as well as how they reflect the properties of their host galaxies, are some of the most pressing questions in astrophysics today. With the discovery of young massive clusters (YMCs) in starburst galaxies in the local Universe, as well as massive intermediate age clusters (IACs, ages $\sim1-10$~Gyr) in galactic merger remnants, the question of how they relate to the ancient (metal poor) GCs has been tackled from a variety of angles. Many studies have adopted extremely old ages for GCs, often leading to the suggestion that they may pre-date the formation of their (eventual) host galaxy \citep[e.g.][]{peebles68, renzini17}.  Other studies invoked conditions unique to the early Universe in order to form GCs \citep[e.g.,][]{Fall_and_Rees_1985, trenti15, kimm2016,  creasey19, madau20}, or at least a sub-set of them (i.e. the low metallicity GCs). In these models, the formation of ancient (metal poor) GCs is fundamentally different from massive clusters forming today, hence there is is little or no link between them.

A contrasting view is that the YMCs, IACs and ancient GCs share the same formation mechanism. The time evolution of the environmental dependencies of such formation mechanism, combined with the effects due to their evolution (i.e.~stellar evolution and mass loss history), would lead to the observed broad spectrum of cluster properties observed today \citep[e.g.~see][]{Krumholz19}. Under this scenario, higher maximum cluster masses and cluster formation efficiencies would be favoured in the extreme conditions of galaxies at the early cosmic times during which GCs formed \citep[][]{elmegreen97, Kruijssen_15}. In this view, the formation of stellar clusters (low or high mass) is tied to star formation in the host galaxy, which in turn is largely controlled by the ambient interstellar medium (ISM) properties of the host at the time of formation \citep{Kruijssen_12}. There is little doubt today that the formation of YMCs and IACs is linked to the environment in which they reside \citep[e.g.,][]{Adamo_Bastian_18}.  The question is whether this also applies to the ancient GCs.

A number of models and simulations of the co-formation and evolution of massive clusters and their host galaxies have adopted this view with some success \citep[e.g.,][]{Kruijssen_15, li2017, Li2018, Choski2018, forbes_review_18, Kim2018, Pfeffer_et_al_18, Adamo2020, lahen20, Ma2020}.  

If cluster formation is a product of general star formation in a galaxy, the cluster will inherit the metallicity of the ISM in which it forms. Hence, we would expect the cluster population to trace out a curve in age-metallicity space, starting with the oldest objects at low metallicity and ending with the youngest objects at higher metallicity \citep{muratov10,Kruijssen_et_al_19b}.  This would mirror the underlying stellar population of the galaxy\footnote{Where recent accretion events generate an additional cluster population that is offset to lower metallicities \citep[e.g.][]{Kruijssen_et_al_19a}.}.  In this view, the metal poor GCs are not a fundamentally different population with an exotic formation channel, rather they are simply part of the early chemical evolution of their host galaxy (e.g. for the Milky Way, see \citealt{Keller2020}).

\begin{figure}
    \centering
	\includegraphics[width=0.47\textwidth]{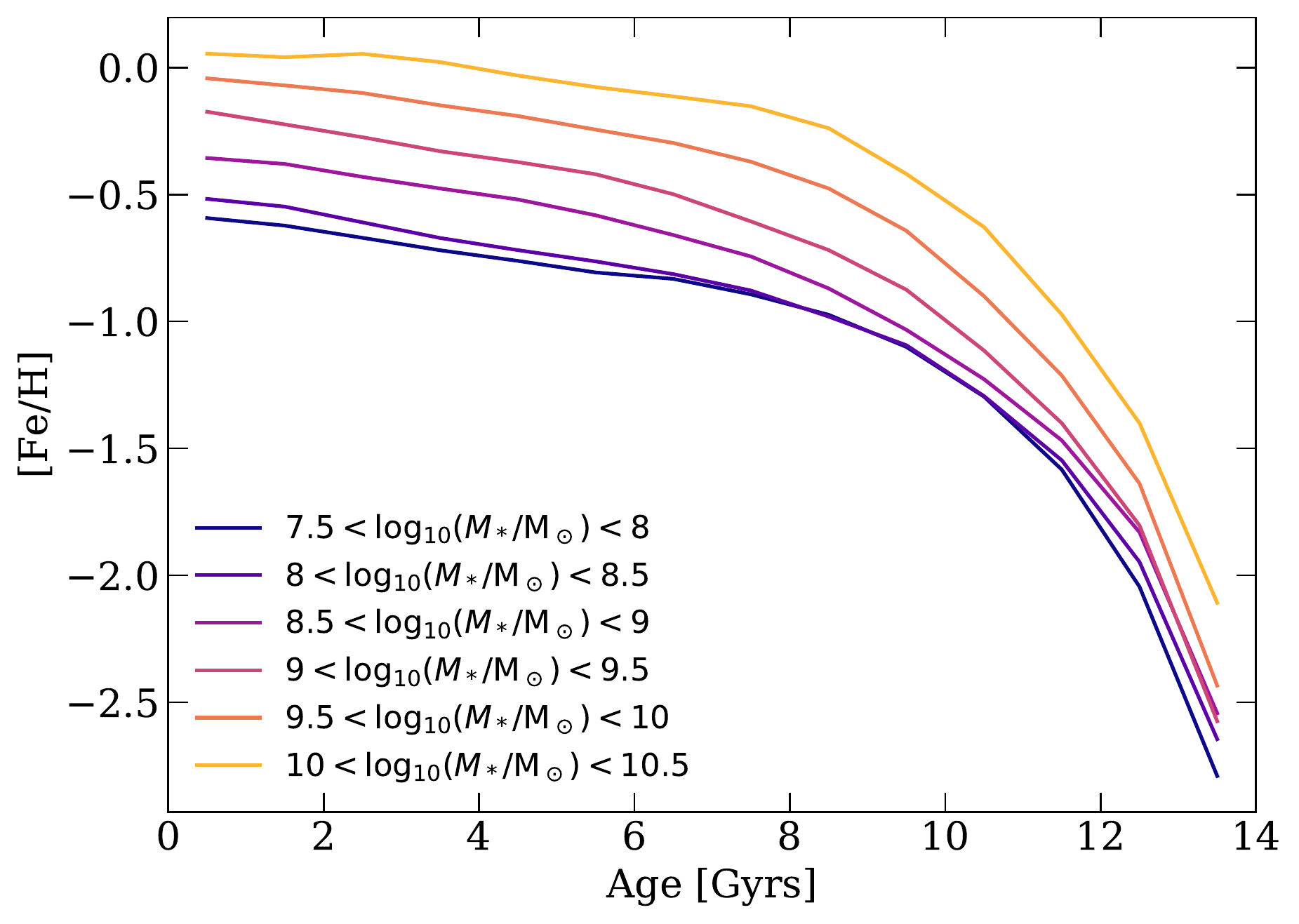}
    \caption{The age-metallicity relation (AMR) for stars in the EAGLE Recal-L025N0752 simulation, binned according to z=0 host galaxy stellar mass.}
    \label{fig:amr_stars}
\end{figure}

A basic result coming from simulations of galaxy assembly is that massive galaxies enrich faster than lower mass galaxies. We show one such result from the EAGLE simulations \citep[][]{Schaye_et_al_15, Crain_et_al_15} in Fig.~\ref{fig:amr_stars}.  Each curve represents the median metallicity of forming stars as a function of age (we will refer to these as age-metallicity relations, or AMR).  High mass galaxies experience extremely rapid enrichment while lower mass dwarfs undergo a more gentle enrichment, leading to an overall lower curve in age-metallicity space towards more metal poor values. This AMR dependence of star formation has also been observed in galaxies in the local Universe \citep[e.g.,][]{Gallazzi_et_al_05,Bellstedt20}.

If massive cluster formation (including YMCs, IACs and the ancient GCs) are tracing the formation of their host galaxy, then we would also expect them to follow such curves \citep[e.g.,][]{muratov10,Kruijssen_et_al_19a}. A number of studies have noted that the Milky Way (MW) GC population has a forked AMR distribution, with a steep component beginning at low metallicities and continuing to the highest metallicity GCs known, along with a branch beginning at [Fe/H]$\sim-1.5$ that extends to lower ages with a much shallower slope \citep[e.g.,][]{Marin-Franch_et_al_09,forbes_bridges10,leaman13}.  This shallower branch has been demonstrated to be made up of accreted clusters \citep[][]{Kruijssen_et_al_19b, Massari_et_al_19, Forbes_20}, and it also appears as a separate branch for metal poor GCs in the $\alpha$-Fe plane \citep[e.g.][]{Horta2020}. It is due to the fact that the accreted galaxies have lower masses (and hence shallower age-metallicity relations) that such a branch exists.  This insight has allowed the unravelling of the assembly history of the Galaxy through its GC population \citep[e.g.,][]{Kruijssen_et_al_20}, and predicted early accretion events which have recently been confirmed with the discovery of its stellar debris \citep{Horta2020b}.

Here, we explore the age-metallicity relation as a function of galaxy mass within the E-MOSAICS simulations \citep{Pfeffer_et_al_18,Kruijssen_et_al_19a}. Specifically, we focus on galaxies with masses similar to that of the Large and Small Magellanic Clouds (LMC and SMC, respectively).  By comparing such dwarf galaxies to MW-like galaxies, the large range in the expected AMRs can be exploited.  The LMC and SMC also host considerable populations of massive clusters, with ages ranging from just a few Myr \citep[e.g.,][]{baumgardt13}, to near the age of the Universe \citep[e.g.,][]{wagner-kaiser17}, and all epochs in between \citep[e.g.,][]{glatt08}.  We focus on the SMC, LMC and MW as the clusters can be studied through resolved CMDs (that reach well below the main sequence turn-off) and spectroscopy of individual stars, giving the tightest constraints on the ages and metallicities of the clusters.  
 
A basic prediction of models that tie GC formation to the star-formation of their host galaxy is that all massive clusters should trace the AMR of the host galaxy.  Alternatively, models that invoke special conditions of the early Universe to form GCs would not predict any variations in the AMR from galaxy-to-galaxy.  If it is found that massive clusters, at all ages, trace a continuous AMR, and that each galaxy (galaxy mass bin) has its own AMR, this would lend strong support to theories that tie together YMCs, IACs and the ancient GCs. 

This paper is organised as follows: in Section~\ref{sec:sims} we discuss the simulations employed in this study, focusing on the sample of L/SMC-mass galaxies in Section~\ref{sec:sims2}, as well as the observational data used for GCs in the LMC and SMC in Section~\ref{sec:obs_data} (see Appendix~\ref{sec:obs_sample} for a more detail description of the observational sample). In Section~\ref{sec:cenvssat} we show the results from the comparison between the AMRs of L/SMC-mass galaxies from the simulations that are classified as centrals or satellites. We then compare the SFHs of the L/SMC-mass galaxies from E-MOSAICS with the SFHs of the LMC and SMC in Section~\ref{sec:sfh}. We discuss the implications of our results in Section~\ref{sec:discussion}, and present our main conclusions in Section~\ref{sec:conclusions}.

\section{Simulations and Observations}\label{sec:sims}

In this paper we explore the star cluster age-metallicity relation (AMR) as a function of galaxy mass and environment. The goal is to test whether GC formation is a natural outcome of star formation in galaxies at any redshift, which is the underlying premise in the E-MOSAICS model (and other recent theoretical and numerical studies - \citealp[e.g. ][]{li2017,Choksi2018,Kim2018,Ma2020}).

To do this we focus on the Large and Small Magellanic Clouds (LMC and SMC, respectively), both of which host relatively large populations of massive ($>10^4$~\msun) clusters that span a wide range of ages, from forming today to $\gtrsim10$~Gyr ago. Additionally, we will exploit the results for Milky Way mass galaxies from \citet{Kruijssen_et_al_19b}. 

\subsection{Simulations and Sample Selection}
\label{sec:sims2}
E-MOSAICS couples the MOSAICS (MOdelling Star cluster population Assembly In Cosmological Simulations) star cluster formation and evolution model (\citealp{Kruijssen_et_al_11,Pfeffer_et_al_18}) to the EAGLE (Evolution and Assembly of GaLaxies and their Environments) galaxy formation and evolution model \citep{Schaye_et_al_15, Crain_et_al_15} in a subgrid fashion. This allows us to follow the formation and evolution of star clusters and their host galaxies simultaneously and in a cosmological context. We will outline the relevant details of the simulations here, for a full description of the models we refer the interested reader to \citet{Pfeffer_et_al_18,Kruijssen_et_al_19a} and Crain et al. (in preparation). 

As a stellar particle is formed, it may form a number of sub-grid star clusters. The number and individual masses of the star clusters formed depends on the local star-forming environment. The fraction of mass that is assigned to cluster formation is regulated by the cluster formation efficiency \citep{Bastian_08}, which in the simulations is related to the local properties of the interstellar medium \citep{Kruijssen_12}. Then, cluster masses are stochastically sampled from a \citet{Schechter_76} initial cluster mass function with an environmentally dependant truncation mass \citep{Reina-Campos_and_Kruijssen_17}. This star cluster population then inherits the age and chemical composition of its parent stellar particle. \citet{Pfeffer_et_al_19a} showed that the star clusters formed in the MOSAICS model reproduce the observed properties of young clusters in the nearby universe.

Star cluster mass loss occurs via stellar evolution (which is calculated for each stellar particle by the EAGLE model, \citealt{Wiersma_Schaye_and_Smith_09}), two-body relaxation and tidal shocks \citep{Kruijssen_et_al_11}. Complete disruption of clusters via dynamical friction is applied in post-processing \citep[see][]{Pfeffer_et_al_18}.  

We use the E-MOSAICS simulation of a large periodic volume ($34.4$ comoving Mpc on a side, L034N1034), which will be presented in detail in Crain et al.~(in preparation; see also \citealt{Bastian_et_al_20}). This is the largest hydrodynamical, cosmological simulation to date that includes star cluster formation and evolution.  The large size of this simulation means that there are hundreds of lower mass galaxies that we can use to make statistical conclusions and comparisons. The simulation adopts the `Recalibrated' EAGLE galaxy formation model and uses a resolution equivalent to the EAGLE Recal-L025N0752 simulation \citep[initial baryon masses of $2.26\times 10^5$~\msun,][]{Schaye_et_al_15}. In this work we consider the fiducial (environmentally-dependent) E-MOSAICS cluster formation model \citep[for discussions of other cluster formation models, see][]{Kruijssen_et_al_19b,Reina-Campos_et_al_19, Pfeffer_et_al_19b, Bastian_et_al_20}.

We select galaxies with masses comparable to the LMC and SMC, namely $1.5\times10^9$~\msun and $4.6 \times 10^8$~\msun\, respectively \citep{mcconnachie12}, allowing for a $\pm 5\times10^8$~\msun\ and $\pm2\times10^{8}$ mass range of the adopted mass of the  LMC and SMC, respectively.  This results in a total number of $200$ LMC and $405$ SMC analogues, of which $120$~LMC and $232$~SMC analogues are central galaxies (i.e., are the dominant halo in their group).  The remaining $80$~LMC and $173$~SMC mass galaxies are satellites of a larger halo. In addition, we also use the 25 MW-mass central galaxy sample presented previously as part of the E-MOSAICS suite \citep{Pfeffer_et_al_18,Kruijssen_et_al_19a}. This sample is unbiased and volume-limited, as it corresponds to  all galaxies with halo masses $7\times10^{11}<M_{200}/M_{\odot}<3\times10^{12}$ from the 25~cMpc$^3$ high-resolution volume Recal-L025N0752 from EAGLE.

We focus on the origin of globular clusters and therefore limit our simulated star cluster population to those more massive than $>10^{4}$~\msun  at $z=0$ throughout this work. We note that we do not make any cuts in age or metallicity of the clusters in the simulations.
As discussed by \citet{Kruijssen_et_al_19a}, the E-MOSAICS simulations of MW-mass galaxies have too low of a disruption rate in the high metallicity inner regions of these galaxies when compared to observational estimates \citep[e.g.,][]{Horta2020c}. Such an effect could plausibly also affect the L/SMC-mass analogues in this work. However, we note that galaxies with stellar masses within the range $10^8$-$10^{9.5}$~\msun\ in the E-MOSAICS simulations have GC specific masses (the fraction of a galaxy's stellar mass within GCs) consistent with observed galaxies \citep{Bastian_et_al_20}, suggesting that under-disruption is less severe within dwarf galaxies.

In Fig.~\ref{fig:amr_obs} we show the resulting AMR relation of massive GCs (namely, >10$^{4}$M$\odot$) for our central L/SMC-mass galaxy samples as orange and purple solid lines, respectively, and the halo mass ($M_{200} \approx 10^{12}$~\msun) selected MW sample \citep{Kruijssen_et_al_19a} in black\footnote{Within this sample there is a factor of five difference between the most and least massive galaxies.}. The shaded regions show the 16$^\mathrm{th}$ and 84$^\mathrm{th}$ percentiles. As expected, the AMRs traced by GCs from different mass galaxies occupy the same AMR position as their star counterparts, displaying similar metallicity values at old ages when compared to more massive galaxies, yet occupying more metal poor positions at younger ages. This result is expected in the E-MOSAICS simulations \citep{Kruijssen_et_al_19a}, and is a result of the longer star formation and chemical enrichment timescales of their smaller mass host.

\subsection{Observations}
\label{sec:obs_data}
There has been a long history of using stellar clusters to trace the age-metallicity distribution of the L/SMC  \citep[e.g.,][]{da_cost91,geisler03,bekki12,parisi14}. These studies have mostly been focused on the chemical evolution of galaxies, implicitly assuming that the chemical evolution of the cluster populations follow that of the underlying stellar field.  We argue here that the ancient GCs of both galaxies are an extension of this trend at early times.

As we are focused on the origin of globular clusters (GCs), in common with the simulations, we limit our sample of LMC and SMC clusters to those with masses $>10^{4}$~\msun.
We note that both the LMC and SMC host clusters with masses $>10^{5}$~\msun, i.e. directly comparable to the ancient GCs in the MW and other galaxies. Our sample of clusters in the L/SMC is given and discussed in Appendix~\ref{sec:obs_sample}.

We note that our sample (for both the LMC and SMC) is incomplete.  This is mainly due to clusters not having accurately measured parameters, specifically their age, metallicity, and/or mass.  However, the sample used is representative of the overall cluster population of both galaxies.  In particular, we note that there are an additional five ancient clusters in the LMC sample which have not had their ages accurately derived (beyond showing that they are likely old ($\gtrsim10$~Gyr, see small triangles in Fig~\ref{fig:amr_obs}). These clusters will be discussed in more detail in \S~\ref{sec:discussion}.

We compare the predicted AMRs to these observations of massive clusters in the LMC and SMC in Fig.~\ref{fig:amr_obs}, which will be discussed in more detail in \S~\ref{sec:sfh} and \S~\ref{sec:discussion}.

\section{Centrals vs. Satellites}
\label{sec:cenvssat}

As the LMC and SMC are currently within the MW's virial radius \citep[e.g.,][]{Guglielmo2014}, they are technically not isolated central galaxies, but are rather satellites of the MW. In this section we look at the differences in the star cluster AMR between centrals and satellites within our simulations.

The main difference between central and satellite galaxies in the simulations is that satellites can have their star formation \citep[e.g.][]{Fillingham_et_al_15, Simpson_et_al_18}, and thus star cluster formation \citep[e.g.,][]{Hughes_et_al_19, Kruijssen_et_al_20}, truncated due to losing their star-forming gas after entering the halo of a larger galaxy.  Since we make a cut on galaxy mass, those satellite galaxies whose star-formation was truncated will have reached a higher metallicity at earlier times than centrals of similar present day mass \citep[see e.g.,][]{Kruijssen_et_al_20}.
This is also observed for galaxies in the Local Group, where dwarf galaxies within the virial radii of the Milky Way and M31 are largely devoid of HI gas \citep[e.g.][]{Grcevich_and_Putman_09, mcconnachie12, Spekkens_et_al_14}.
Hence, for a given galaxy mass at the present day, satellite galaxies will need to build-up their stellar mass rapidly before they are truncated.  Centrals, on the other hand can build up their mass over the full cosmic history \citep[e.g.,][]{Mistani_et_al_16}, since their star/cluster formation is typically not truncated. From this mass assembly bias, satellite galaxies (within the selected mass range) are expected to have steeper AMRs than centrals (i.e., they enrich more rapidly than centrals).

\begin{figure}
    \centering
	\includegraphics[width=0.47\textwidth]{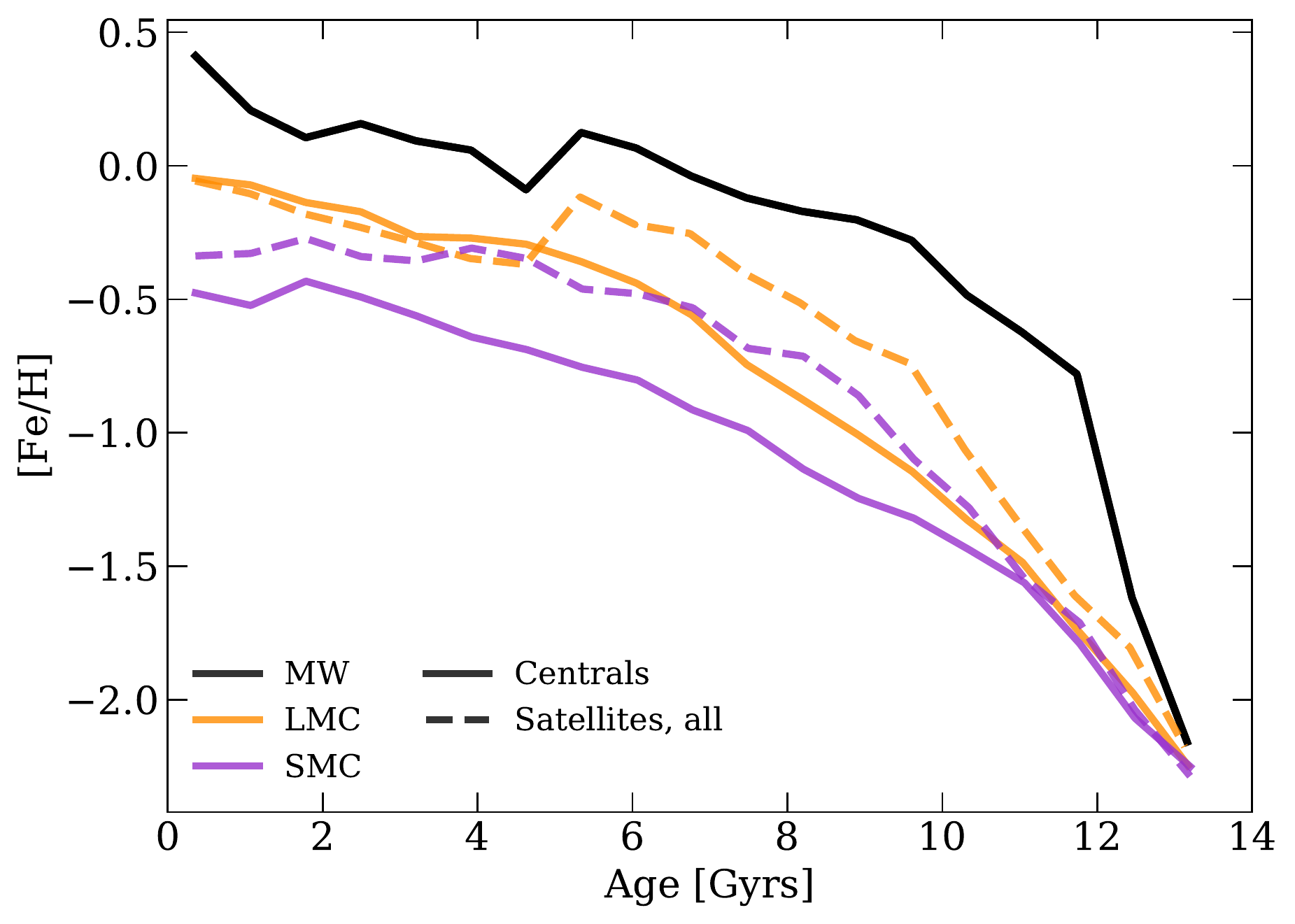}
    \caption{The median star cluster AMR relation for LMC (orange), SMC (purple), and Milky Way-mass (black) galaxies in the simulations. Solid lines show the relation for centrals while dashed lines are for satellite galaxies.  The relations for satellites lie systematically above that for centrals.  This is due to assembly bias at the selection of galaxies of a specific mass range at $z=0$.}
    \label{fig:sat1}
\end{figure}

\begin{figure}
    \centering
	\includegraphics[width=0.47\textwidth]{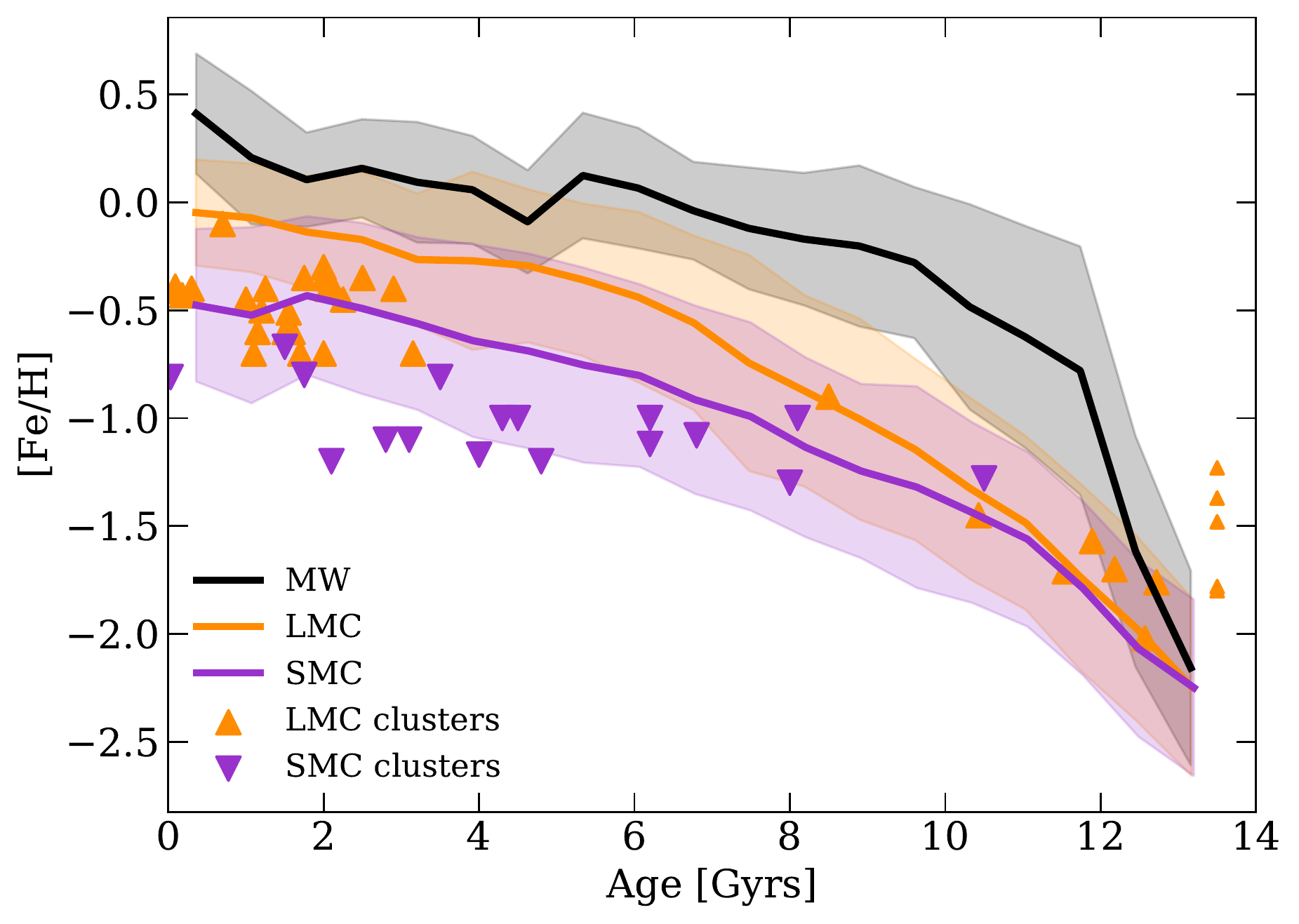}
    \caption{The AMR for massive clusters in the LMC (orange triangles) and SMC (purple, upside-down triangles) as well as the median and $1\sigma$ AMRs for massive clusters ($>10^{4}$~M$\odot$) in the simulated central S/LMC-mass galaxies (purple and orange, respectively), and for Milky Way-mass (black) galaxies. The MW AMR was taken from \citet{Kruijssen_et_al_19b}, and only includes GCs with massess $>10^{5}$~M$_\odot$. The solid lines show the median relation while the shaded regions denote the 16$^{\rm th}$ and 84$^{\rm th}$ percentiles of each distribution. The highest metallicity ancient LMC cluster, NGC 1898, should provide an excellent tests of the models, as its age should be $\lesssim11$~Gyr. For the LMC, five ancient clusters without detailed age fitting are shown as small triangles at an arbitrary age of 13.5~Gyr, in order to highlight their metallicities. A small offset in metallicity may be present due to systematic uncertainties in nucleosynthetic yields in simulations \citep[which are uncertain at a factor of 2 level, e.g.][]{Wiersma_et_al_09} or an offset in the adopted L/SMC mass compared to the true value.}
    \label{fig:amr_obs}
\end{figure}

In Fig~\ref{fig:sat1} we show the AMR of massive clusters for our L/SMC-mass galaxy samples, both for centrals (solid lines, the same as in Fig~\ref{fig:amr_obs}) and for satellites (dashed lines). We note that the AMR for the satellite galaxies is systematically above that for centrals as expected from mass assembly bias.

\begin{figure}
    \centering
	\includegraphics[width=0.47\textwidth]{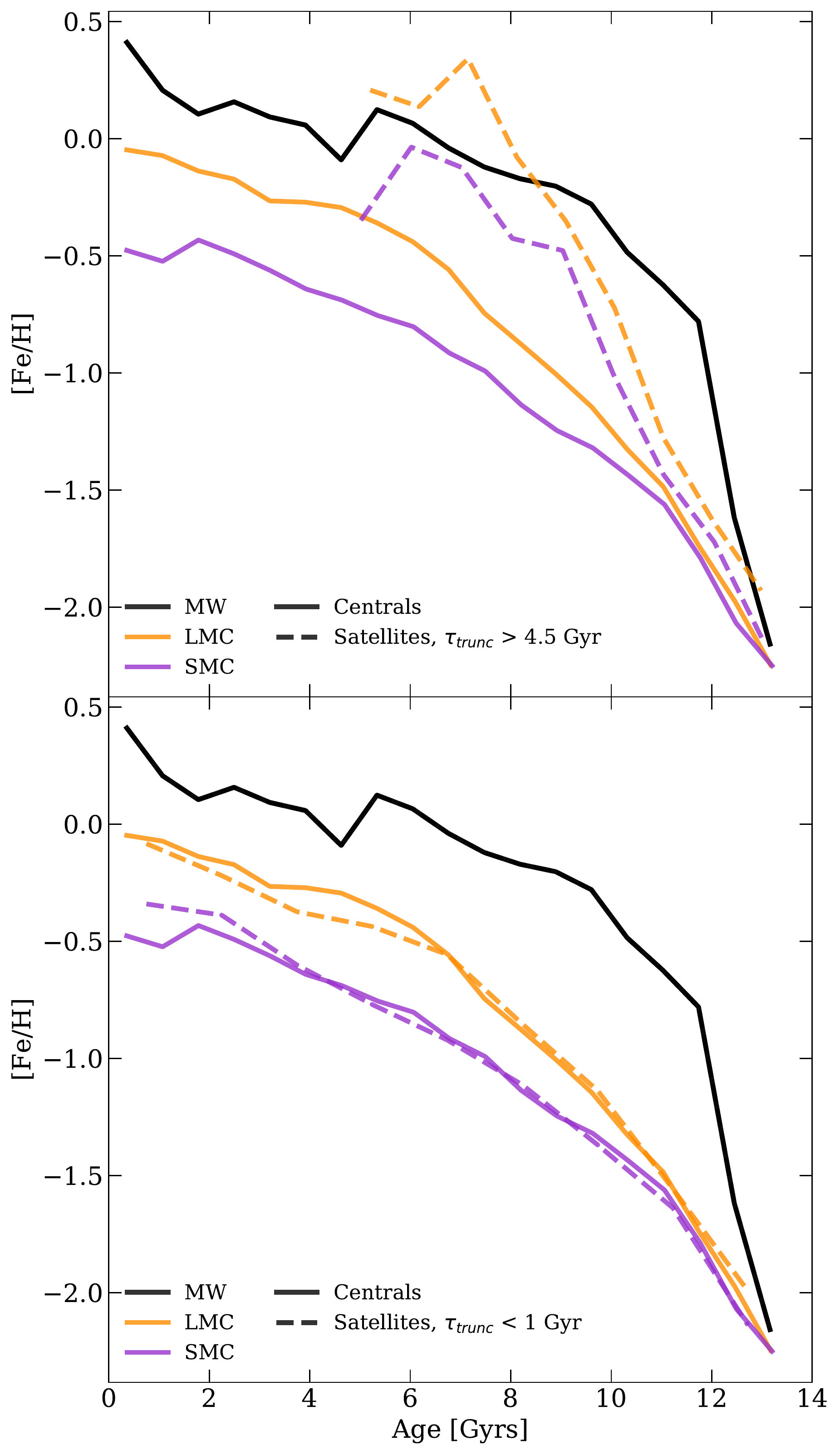} \caption{The same as Fig.~\ref{fig:sat1}, but selecting galaxies that have had their cluster formation truncated at time $\tau_{\rm trunc}$. The upper panel shows the AMR for satellites with relatively early truncation of their cluster formation ($\tau_{\rm trunc}$~$>4.5$~Gyr) while the bottom panels includes simulated galaxies that have continued to form clusters until the past Gyr. We find that satellites that have not had their star/cluster formation truncated until recently (or at all) are the best representative of the L/SMC system due to their recent infall time into the larger host halo. }
	\label{fig:sat2}
\end{figure}

We investigate the origin of this offset in Fig.~\ref{fig:sat2}, where we split the satellite galaxy populations into sub-samples based on whether or not they have had their cluster formation truncated. Although accretion onto a larger host galaxy is the most likely cause for such truncation, we note that we do not categorise the galaxies based on infall time, but rather when the galaxy stopped forming clusters. In the top panel, we only include satellite galaxies that have had their cluster formation truncated at time $\tau_{\rm trunc} >4.5$~Gyr. Here, the offset is much more pronounced and is driven by our selection criteria of a fixed stellar mass at $z=0$. Galaxies that truncated their star/cluster formation early, must have had a higher mass at $\tau_{\rm trunc}>4.5$~Gyr than galaxies that continued to form stars/clusters, in order to have the same mass at $z=0$. 

However, we note that both the LMC and SMC have continued to form massive clusters within the past Gyr, i.e. their star/cluster formation has not yet been truncated by their infall into the Milky Way halo (see Appendix~\ref{sec:obs_sample}). This is expected given their (probable) recent infall \citep[$\lesssim 4$~Gyr,][]{Rocha_et_al_12, Patel_et_al_17} and the long quenching timescale for L/SMC-mass galaxies \citep[which is expected to occur via starvation,][]{Wheeler_et_al_14, Fillingham_et_al_15}.
In the bottom panel of Fig.~\ref{fig:sat2} we show a sub-sample of satellite galaxies that have continued to form massive clusters until (at least) the past Gyr, in order to match the LMC and SMC.  Here we see that there is no significant offset between the AMR of the satellites and centrals. This result is in agreement with findings by \citet{Garrison-Kimmel2019}, who studied the SFHs of dwarf galaxies with a range of masses in the FIRE-2 simulations \citep{Hopkins2018}. Specifically, these authors compared the SFHs of satellite galaxies and centrals, and found that for L/SMC-mass galaxies (see \S~\ref{sec:sfh}), the SFH of centrals near MW mass galaxies were most alike the SFH of the LMC and SMC galaxies.

We conclude that the AMR is the same for central galaxies and satellites that match the cluster formation duration of the LMC and SMC. Therefore, in the rest of this work we use the simulated central galaxies as representations of the LMC and SMC \footnote{We also check if the results obtained using the satellite galaxies which have not had their star cluster formation truncated are consistent with the centrals, and find that these are in agreement.}.

\section{Star Formation Histories}
\defcitealias{weisz_et_al13}{W13}
\label{sec:sfh}

\begin{figure*}
    \centering
	\includegraphics[width=\textwidth]{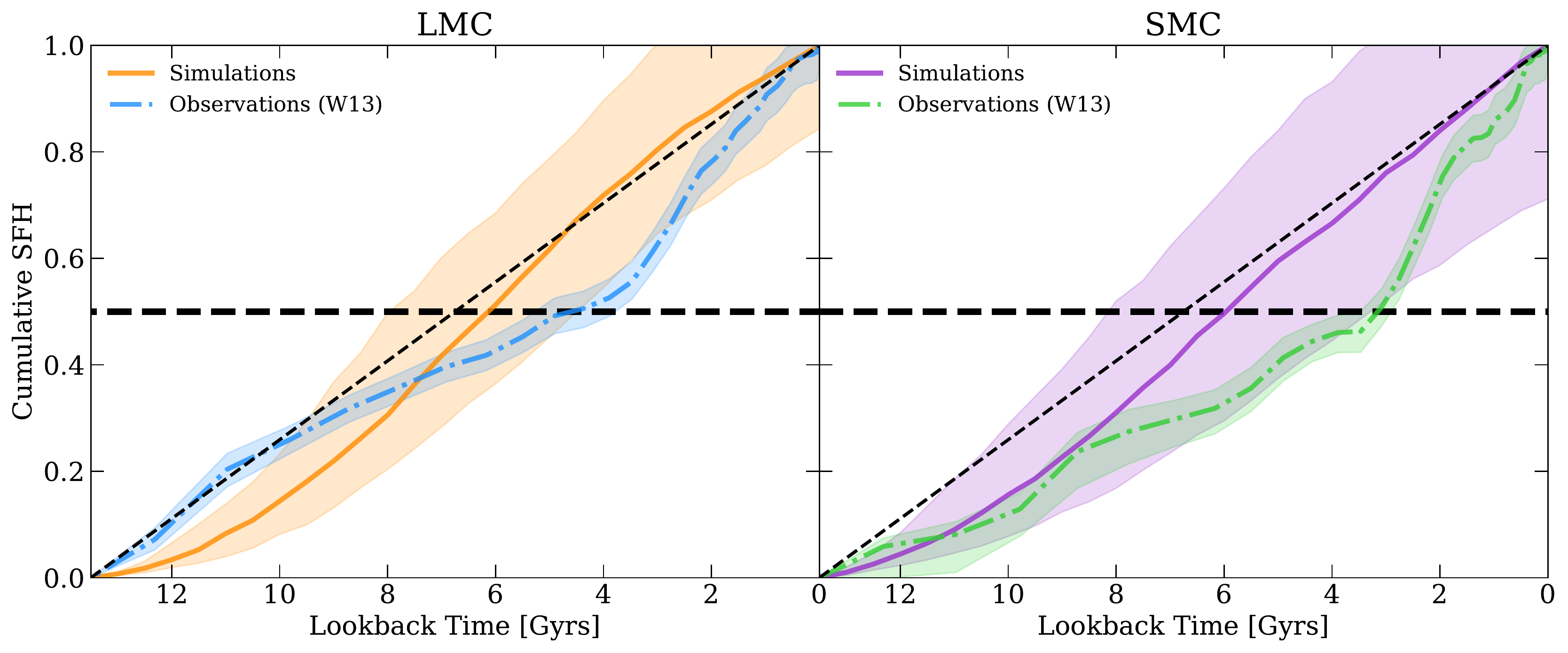}
    \caption{Cumulative SFHs as a function of lookback time of L/SMC-mass central galaxies from the simulations (orange/purple) and the observations from \citetalias{weisz_et_al13} (blue/green), where the straight line signifies the respective median values and the shaded region the 16$^{\rm th}$ and 84$^{\rm th}$ percentiles. Horizontal dashed line signifies the point the systems built $50$~per cent of their mass, whereas the diagonal dashed line shows the linear correlation. The predictions from the simulations are in agreement with the observations within 1 $\sigma$. The LMC has a very different SFH to the SMC
    , building $\sim$ 20~per cent of mass within the first $\sim$ 2 Gyr. Conversely, the SMC is consistent with having approximately no SF until $\sim$ 11 Gyrs ago. Furthermore, the LMC appears to obtain half of its present day mass $\sim$ 1 Gyr before the SMC. From this moment onwards, both galaxies then show a similar increase in SF until reaching their present day mass.}
    \label{fig:sfh_mcs}
\end{figure*}

Due to their proximity, the LMC and SMC offer a unique opportunity to study their evolution, as their (field) stellar populations can be resolved down to and beyond the ancient main sequence turnoff.  This has allowed the reconstruction of each galaxy's star formation history (SFH) in a detail not possible for nearly all other stellar systems.  This allows us to directly compare the formation of the field star component of the galaxies to that of their star cluster populations.

For the observational comparison we use the SFHs as estimated by \citet[hereafter \citetalias{weisz_et_al13}]{weisz_et_al13}.  As noted by the authors, the LMC and SMC have undergone very different assembly histories, which we may expect to be reflected in their cluster populations.

In Fig.~\ref{fig:sfh_mcs} we show the median (plus the 16$^{\rm th}$ and 84$^{\rm th}$ percentiles) SFHs of the simulated central LMC (left panel) and SMC (right panel) mass galaxies. We also show the observed distribution from \citetalias{weisz_et_al13}.  Overall, the simulations (and the scatter) broadly reproduce the SFHs of the LMC and SMC, with the notable exception of the LMC at the oldest ages ($>10$~Gyr ago). 

In order to find simulated galaxies that best match the observed SFH of the LMC and SMC, we extract the SFH for each of the simulated galaxies in the sample (both centrals and satellites) and compared them to the observations by calculating a $\chi^{2}$ value. The most similar L/SMC analogues are those for which the resulting $\chi^{2}$ value is smallest, and are shown in Fig~\ref{fig:lmc_analogues} and Fig~\ref{fig:smc_analogues}, respectively.

As noted, none of our simulated galaxies reproduce the ancient LMC SFH, hence we restrict the comparison on lookback times within $\lesssim 5$ Gyr. In each of the panels we show the time of formation of clusters with masses $>5\times10^{4}$~\msun\ as dots along the cumulative track.  The same is shown for the observations, where we only show the 'high confidence' sample (i.e.~those clusters with accurate mass, age and metallicity measurements).

For the SMC-analogues, we find a number of simulated galaxies that closely match the observed SFH of the SMC. In the following subsections, we break up the SFH into separate epochs in order to more directly compare theory and observations. We find that the AMR for GCs in the satellite L/SMC-mass galaxies from Fig~\ref{fig:lmc_analogues} and Fig~\ref{fig:smc_analogues} follow the same relation as the centrals from Fig~\ref{fig:amr_obs}.

Ideally, we would like to quantitatively compare the observed star and cluster formation histories of the LMC and SMC clusters with that of our simulations.  Unfortunately, this is currently not possible, as we are limited by the incompleteness of the observed cluster catalogues.  For example, the LMC hosts a number of ancient GCs that are not included in our study (see Table~\ref{tab:obs_sample}) as they lack detailed age information (as well as [$\alpha$/Fe] measurements).  Similarly, it is likely that the current catalogues are incomplete, especially at masses below >$5\times10^{4}~\msun$, at intermediate ages.
 Below we qualitatively compare the star and cluster formation histories of both the LMC and SMC and compare them to expectations from our simulations.

\subsection{The Ancient SFHs ($>10$Gyr)}

According to \citetalias{weisz_et_al13}, by 10~Gyr ago, the LMC had built up $\sim27$\% of its present stellar mass (i.e., $\sim4\times10^{8}$~\msun).  In comparison, by the same age the SMC had only assembled $\sim12$\% of its $z=0$ mass (i.e., $\sim5\times10^{7}$~\msun).  This difference is compounded by the respective masses of the LMC and SMC, with the LMC being $\sim3$~times more massive at the present time. As such, we expect that the LMC had a mean star-formation rate at early times (up until $10$~Gyr ago) that was $\sim6$~times higher than the SMC.

This rapid early assembly of the LMC, relative to the SMC, would be expected to result in the formation of a larger cluster population in the LMC.  This is borne out in the observations, where the LMC hosts $11$ GCs older than $10$~Gyr ago, while the SMC only hosts 1 (NGC~121).  In fact, the SFH of the SMC is actually consistent with little or no star formation happening for lookback times >11~Gyr ago. 

More massive cluster formation at early times is also predicted for the simulated LMC-mass galaxies. We find that six out of the ten LMC-mass analogues form $\geq$5 massive clusters (i.e., 5$\times$10$^{4}$~M$_\odot$) before a lookback time of $\sim$10~Gyr (see Fig.~\ref{fig:lmc_analogues}). Conversely, our results reveal that only one SMC-mass analogue galaxy from Fig.~\ref{fig:smc_analogues} (SMC\,52) forms as many massive clusters in the same amount of time, matching the observed number.

\subsection{Intermediate Ages ($3-10$~Gyr)}

The cluster population of the LMC has a notable 'age-gap', which lasts from $\sim3-4$~Gyr to $\sim9$~Gyr.  We note that the SFH of the LMC is quite low during this period (as seen by the shallow slope in Fig.~\ref{fig:sfh_mcs}).  This reduced star-formation activity would be expected to lead to a low cluster formation rate and potentially to a lower truncation mass in the cluster initial mass function \citep[e.g.,][]{Reina-Campos_and_Kruijssen_17}. This interpretation is supported by the fact that simulated LMC-mass analogues with low SFRs (for example LMC\,131, LMC\,158, and LMC\,191) generally have low gas pressures and correspondingly low GC initial mass function truncation masses \citep{Pfeffer_et_al_19b}. Hence, we would expect few or no massive clusters during this age range, in agreement with the observed 'age-gap' in the LMC cluster population. This hypothesis is corroborated by our predictions from the simulations (see Fig.~\ref{fig:lmc_analogues}), where we find that LMC-mass analogues which show a similar SFH as the LMC in this age range (LMC\,26, LMC\,131, LMC\,158, and LMC\,191) show a similar lack of massive cluster formation during that time, displaying the observed LMC ``age gap''.

On the other hand, the SMC hosts a relatively large population of clusters with ages between $4$ and $8$~Gyr, including at least five with masses in excess of $5\times10^{4}$~\msun.  These clusters represent the increased SFR and cluster formation rate (CFR) of the SMC, as it built up its stellar mass after a relatively slow start. When comparing the observed SMC GC AMR to the predictions from the simulations in Fig. \ref{fig:smc_analogues}, we find that there are only two SMC-mass galaxy analogues (SMC\,27 and SMC\,71) that show a similar massive cluster formation as the observations  within this age range, forming approximately 3--4 clusters. Therefore, our results suggest that the increased SFH of the SMC within this age gap, which lead to the formation of the five massive clusters observed, is not typical of SMC-mass analogue galaxies. 

\subsection{Young ages ($<3$~Gyr)}

\begin{figure*}
    \centering
	\includegraphics[width=\textwidth]{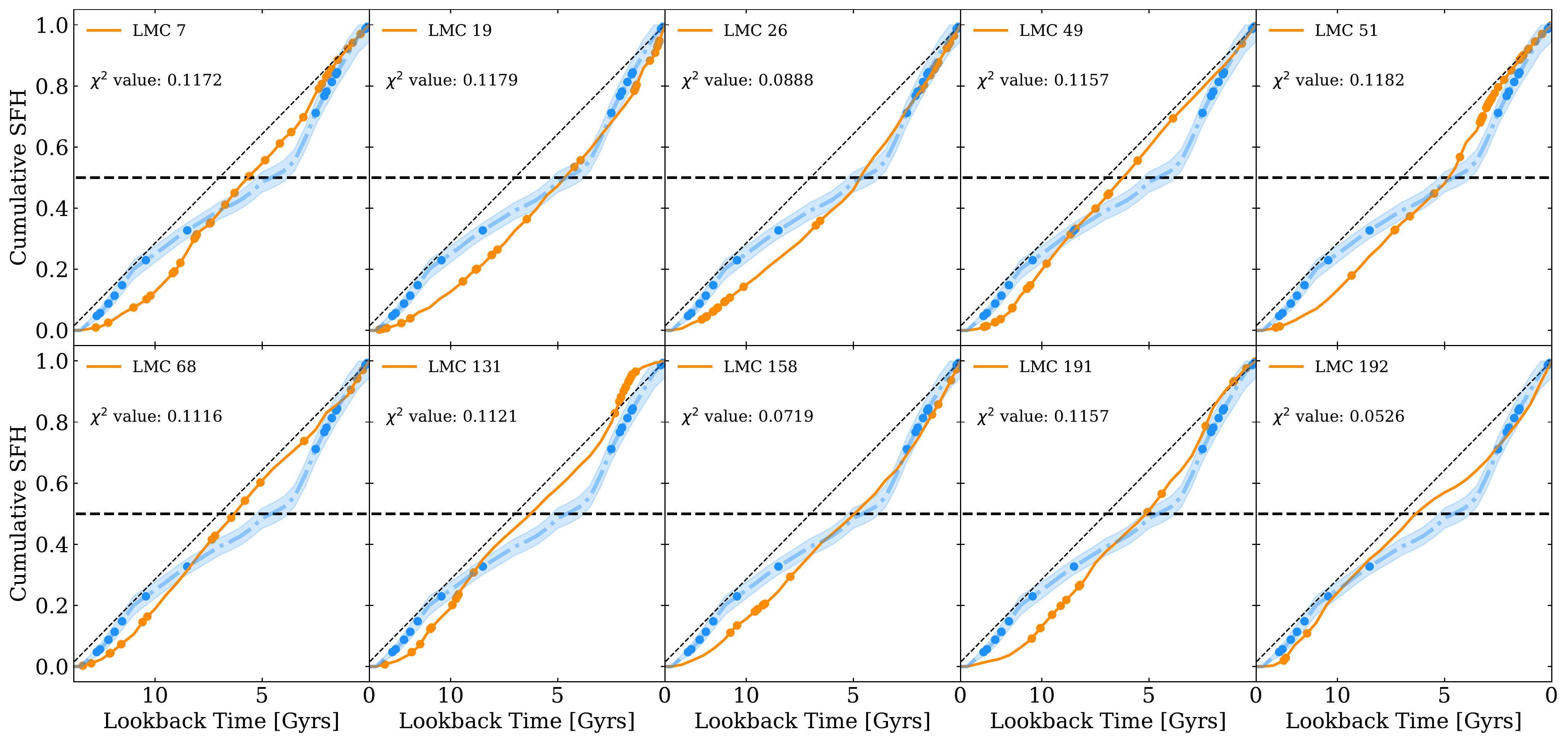}
    \caption{Cumulative SFHs as a function of lookback time for the ten most resembling LMC analogues (orange), selected on a $\chi^{2}$ value merit, are compared with the \citetalias{weisz_et_al13} SFH from Fig. \ref{fig:sfh_mcs} (blue). The dots along the lines represent the massive GCs (namely, $> 5 \times 10^{4}$ M$\odot$) formed in each galaxy, respectively. We find no LMC analogue in the simulations which shows a similar high SFH at high lookback times. However, below a lookback time of $\sim$ 6 Gyr, a number of simulated galaxies are similar to the observed LMC distribution. The time at which 50~per cent of the stellar mass of the galaxy has formed is shown as a horizontal dashed line and a conference constant SFH is shown as a dotted line.
    }
    \label{fig:lmc_analogues}
\end{figure*}

\begin{figure*}
    \centering
	\includegraphics[width=\textwidth]{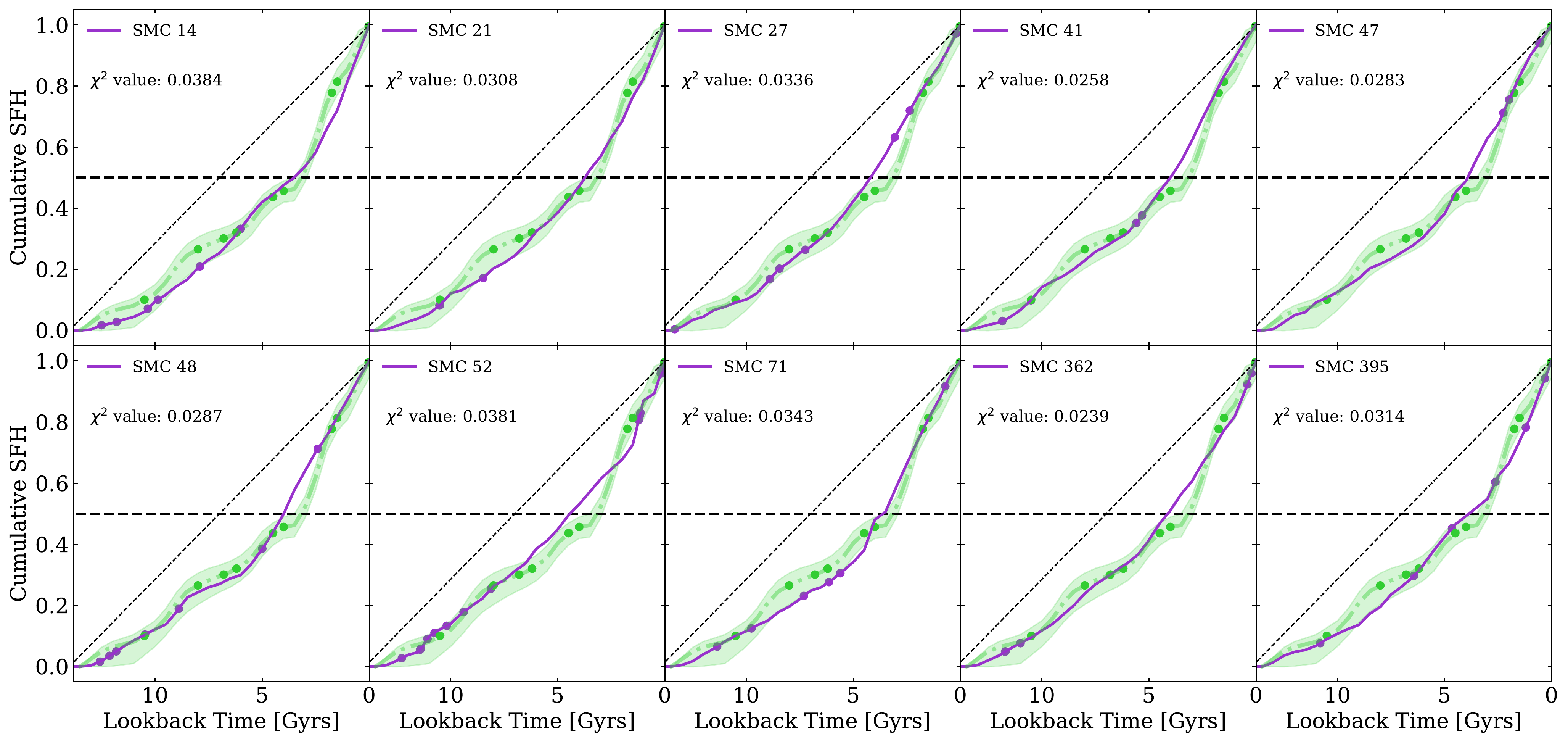} \caption{The same as Fig~\ref{fig:lmc_analogues} but now for the SMC analogues (purple) and the SMC SFH from \citetalias{weisz_et_al13} (green). The SFH predictions for SMC analogues appear to be in good agreement with the observations, particularly for SMC~48, while the GC populations in SMC is closely matched by SMC~27. 
	}
	\label{fig:smc_analogues}
\end{figure*}

Both the LMC and SMC underwent a rapid buildup of their stellar masses during the past $\sim3$~Gyr, with the LMC forming $\sim35$~per cent of its stars and the SMC forming nearly $50$~per cent.  As such, we may expect to see relatively large cluster populations in both galaxies during this epoch.  Again, this is borne out by observations.  The LMC hosts at least 17 clusters more massive than $10^4$~\msun\ and with ages between $1-2$~Gyr \citep[e.g.,][]{goudfrooij14}, and at least $9$ more younger than $1$~Gyr \citep[][]{baumgardt13}. The more abundant population of massive clusters at young ages is also predicted for five out of ten of the LMC-mass analogue galaxies compared in Fig.~\ref{fig:lmc_analogues}. Specifically, we find that these five galaxies (LMC\,7, LMC\,19, LMC\,26, LMC\,51, and LMC\,131)  form five or more massive clusters, matching the high number of young massive clusters that is seen in the observations. Therefore, the majority of the simulated LMC-mass analogues predict this enhanced massive cluster formation at young ages, as expected given their enhanced SFR in this age range. 

The SMC also hosts a population of massive clusters with ages $<3$~Gyr (although less abundant), including the massive ($1.4\times10^{5}$~\msun) cluster, NGC~419 \citep[][]{kamann18}. We find that only four of our ten SMC-mass galaxy analogues have a comparable number of massive clusters at young ages ($\sim$3). 

\section{Discussion}
\label{sec:discussion}
As shown in Fig. \ref{fig:amr_stars}, the AMR normalisation of a galaxy correlates with its stellar mass, whereby less massive galaxies show a lower metallicity increase with decreasing age, than more massive galaxies.  This mass-AMR connection of galactic field populations has been attributed to the star formation history and chemical enrichment of the host galaxy \citep[e.g.][]{Gallazzi_et_al_05}. When attempting to link GCs to this scenario, a series of fundamental questions emerge, such as: how do massive clusters form; do massive clusters follow the AMR of the galaxy they are associated with; and if so, can we use massive clusters to trace the star formation of their host galaxy. In this section, we set out to tackle these pivotal questions by comparing the predicted AMRs of LMC and SMC mass galaxies from the E-MOSAICS simulations with observational results of massive clusters in the Magellanic Clouds.

The origin of these AMR trends in the stellar populations, as a function of galaxy mass, are well known within the context of galaxy evolution.  Higher mass galaxies undergo more rapid enrichment, as they are able to retain the ejecta of high-mass stars and SNe, allowing for a rapid enrichment of the ISM. Lower mass galaxies build up their stellar mass more slowly over time, and lose a larger fraction of the ejecta of stars/SNe, meaning that they enhance their overall metallicities more slowly. Hence, a basic prediction of any model that correlates the formation of massive stellar clusters with the star-formation of a host galaxy, is that the clusters will span a range of ages, with lower mass galaxies hosting younger clusters (at fixed metallicity) compared to massive galaxies.  

The galaxy mass-dependence on the AMR of GC populations has been highlighted in previous works. While the GC population of the Milky Way is quite old on average ($>11$~Gyr), it has been long noted that there are two branches of the MW AMR \citep[e.g.,][]{Marin-Franch_et_al_09, forbes_bridges10, leaman13}. The steep branch represents the in-situ component, i.e. GCs forming in the relatively massive MW progenitor, while the shallower branch represents the GCs accreted from lower mass dwarf galaxies \citep[e.g.,][]{Kruijssen_et_al_19a}. It is this separation that has allowed the assembly history of the MW to be inferred from the AMR relation of its GC population \citep[e.g.,][]{Kruijssen_et_al_19b,Kruijssen_et_al_20}.

As the LMC likely had a mass comparable to the satellites accreted during some of the main accretion events of the MW during its early evolution ($M_{*}\sim10^{8}$~\msun, \citealt{Kruijssen_et_al_20}), we would expect LMC clusters to display an AMR very similar to the MW 'accreted' branch (\citealp{leaman13}, made up mainly of Gaia-Enceladus/Sausage, Sequoia, Sagittarius and Kraken GCs - e.g., \citealt{Massari_et_al_19}), at early times (i.e. before these galaxies were accreted). As the SMC has a lower mass, its AMR is expected to be shallower.

Figure~\ref{fig:amr_obs} displays the AMR for massive clusters (namely, $M_{*}$ > 10$^{4}$~\msun) formed in the simulations in the SMC, LMC and MW analogue galaxy populations as purple, orange and black lines, respectively. The shaded regions denote the 16$^{\rm th}$ and 84$^{\rm th}$ percentiles for each relation. It is evident from Fig.~\ref{fig:amr_obs} that at metallicities below [Fe/H] < --1.5, the AMRs of these three distinct populations of galaxies converge --largely independent of environment-- and should be very old (Age > 10 Gyr). Conversely, despite the MW, LMC and SMC simulations all showing an increase in metallicity with decreasing age, at metallicities above [Fe/H]$=-1.5$, the median ages of the GC populations diverge strongly. This result corroborates that massive clusters follow the AMR of the galaxy they are associated with. At [Fe/H]$=-1.5$, the median age of GCs in L/SMC-mass galaxies is expected to be $\sim2$~Gyr younger than in MW-mass galaxies.  By a metallicity of [Fe/H]$=-1.0$ the median age difference, relative to MW-mass galaxies, has grown to $\sim3$ and $\sim4$~Gyr for LMC and SMC-mass galaxies, respectively.

Figure~\ref{fig:amr_obs} shows the comparison between the predictions from the simulations with observational data for clusters more massive than $M_{*}$ > 10$^{4}$ M$\odot$ in the LMC and SMC galaxies\footnote{We note that a similar result is obtained when including lower mass clusters, however, our goal is to compare the massive cluster populations between galaxies to test whether GCs, intermediate age, and young massive clusters share a common formation/evolution mechanism.}. The predictions from the simulations are generally in good agreement with the observational data, and follow the predicted AMR for their corresponding galaxy mass within 1 $\sigma$. A small offset in metallicity may be present due to systematic uncertainties in nucleosynthetic yields in simulations \citep[which are uncertain at a factor of 2 level, e.g.][]{Wiersma_et_al_09} or an offset in the adopted L/SMC mass compared to the true value. 

The comparison between the observed and predicted AMRs shows that at metallicities below [Fe/H] $\sim-1.5$ the observed/predicted clusters from the LMC and LMC occupy the same AMR locus as the MW predictions, suggesting that the majority of metal poor GCs should be old ($>11$~Gyr).  For the LMC, this is in agreement with the results of \citet{wagner-kaiser17} who found that the metal poor GCs have the same age (within the uncertainties) as MW GCs in the same metallicity range.

We draw particular attention to the more metal rich LMC ancient GCs that have not been subjected to precision age-dating (shown in Fig.~\ref{fig:amr_obs} as smaller triangles at the arbitrary age 13.5 Gyr). The two most metal rich clusters, NGC~1898 and NGC~2019, have iron abundances of [Fe/H]$=-1.2\pm0.2$ and $-1.3\pm0.2$, respectively \citep{johnson06,piatti18}. A direct prediction of the simulations presented here is that, with these metallicities, both clusters should be relatively young, with ages $\sim10\pm$1.5~Gyr (when adopting an age scale of $\sim12$~Gyr for the lower metallicity clusters). This is significantly younger than the metal poor clusters studied previously \citep{wagner-kaiser17}. 

Due to the low number statistics associated with the one SMC GC with an age $>10$~Gyr, we are unable to make the same claim (i.e. that ancient GCs should be metal-poor) for the observed old SMC clusters. However, we note that the star-formation histories of the LMC and SMC were markedly different at early times, with the SMC showing a much lower star-formation rate than the LMC \citep[][]{weisz_et_al13}.  In the E-MOSAICS model, the result of these different SFHsf is that we would expect the LMC to have many more ancient GCs than the SMC \citep[e.g.,][]{Pfeffer_et_al_18}, in agreement with the observations.

At younger ages, the differences in metallicity between the SMC, LMC and MW become even more pronounced. The observed AMR of massive SMC clusters is clearly shallower than that of the LMC clusters, which in turn is shallower than that of the MW.  While our sample of massive LMC and SMC clusters is not complete (due to missing resolved CMDs and/or resolved spectroscopy of stellar members to determine each clusters abundances) the median ages of the SMC, LMC and MW GCs are $7$, $9$, and $12$~Gyr, respectively.

Alternatively, for models that adopt some special conditions in the early universe in order to form GCs, e.g., dark matter mini-haloes \citep[e.g.,][]{trenti15} or high velocity interactions \citep[e.g.,][]{madau20}, no such AMR would be expected.  In these models, once the unique environment needed to form GCs is gone, no further GCs can form.  The MW GC population is bi-modal, with peaks at [Fe/H]$\sim-0.7$ and $-1.3$, and a minimum between the populations at [Fe/H]$\sim-1.0$.  This has led some investigators to propose that the metal poor GCs formed differently than the metal rich GCs, with the former forming under some unique early Universe conditions. We note, however, that the SMC is an excellent litmus test for this class of theories, as it hosts a number of massive clusters with [Fe/H]$<-1.0$ but with ages that extend to $\sim4$~Gyr ago, spanning an age range of $>6$~Gyr. This demonstrates that metal-poor GCs formed until quite recently in dwarf galaxies.

\section{Conclusions}
\label{sec:conclusions}
In summary, the results presented in this study reveal that the formation of the  ancient globular clusters (GC), intermediate-age (IAC), and young-massive clusters (YMC) is interwoven, with massive clusters forming whenever the gas conditions produce intense bursts of star formation across cosmic time and galactic environment.

By comparing the predicted GC age-metallicity relations (AMR) of L/SMC-mass galaxies from the E-MOSAICS 34~cMpc$^3$ volume (Crain et al. in prep.) with those determined observationally, we have shown that there is good agreement between theory and observations. In particular, we have shown that at old ages, the AMR of different mass galaxies converges.

As galaxies evolve and undergo different star formation histories and chemical enrichment, GCs continue to form, and adopt the properties of their host system, yielding the spread of predicted/observed cluster metallicity values at fixed age. Our results thus suggest that the AMR is a powerful tool for understanding massive cluster formation and the formation and evolution of galaxies.

If the formation of massive clusters is indeed a natural outcome of the star-formation process \citep[e.g.,][]{Kruijssen_15,Pfeffer_et_al_18}, implied here by the correlation between the SFH and cluster formation history in the Magellanic Clouds, it would offer a powerful probe into the assembly history of galaxies. This has been shown for the Milky Way \citep[][]{Kruijssen_et_al_19b,Kruijssen_et_al_20,Forbes_20}, and for the Magellanic Clouds in the present paper. Extending this type of work to galaxies at larger distances where resolved stellar studies are in general no longer possible requires integrated light studies. There have been promising steps in this direction, using the AMR to constrain the assembly history of nearby galaxies, such as M33 \citep[][]{beasley15} and more massive early type galaxies \citep[e.g.][]{Usher_et_al_19}.

We have also highlighted the caveats introduced by a truncation of star and cluster formation within satellite galaxies. When selecting by present day galaxy mass, satellite galaxies that have had their SF truncated at early times, will have had to build up their stellar mass much more rapidly than galaxies that were able to form stars continuously over cosmic time. The truncated satellite galaxies will follow steeper AMRs, effectively behaving as higher mass galaxies (as they would have continued to grow if star-formation had not been truncated). This has important effects on the GC population of those galaxies \citep[e.g.][]{Mistani_et_al_16}. 

One common feature shared between metal-poor and metal-rich GCs, also over a wide range of ages (from $\sim2 - 13$~Gyr), is that they host light-element abundance spreads within them. The origin of these abundance variations is still unknown \citep[e.g.,][]{bastian18}, but their ubiquity points to a common formation mechanism across all massive clusters \citep[e.g.,][]{Martocchia19,saracino19}. This continuity, also seen in the AMR of massive clusters in both galaxies presented here, lends support to the underlying assumption of the E-MOSAICS simulations (as well as others within the literature) that GCs can form throughout cosmic history, and that the YMCs and IACs share the same formation and evolutionary mechanisms as the classical, ancient GCs.

\section*{Acknowledgements}
We thank the many people around the world whose hard work fighting the ongoing COVID-19 pandemic has made it possible for us to remain safe and healthy for the past several months.
We also warmly thank the organisers of the KITP20Clusters conference during which much of this paper was developed. DH thanks Sue, Alex and Debra for always being there, and thanks Ricardo P. Schiavon for being understanding during the current uncertain times. DH acknowledges an STFC studentship. NB, MH, and JP gratefully acknowledge financial support from the European Research Council (ERC-CoG-646928, Multi-Pop). NB and RAC  acknowledge support from the Royal Society (University Research Fellowship). JMDK and MRC gratefully acknowledge funding from the European Research Council (ERC) under the European Union’s Horizon 2020 research and innovation programme via the ERC Starting Grant MUSTANG (grant agreement number 714907). JMDK gratefully acknowledges funding from the Deutsche Forschungsge-meinschaft (DFG, German Research Foundation) through an Emmy Noether Research Group (grant number KR4801/1-1) and the DFG Sachbeihilfe (grant number KR4801/2-1). MRC is supported by a Fellowship from the International Max Planck Research School for Astronomy and Cosmic Physics at the University of Heidelberg (IMPRS-HD). This research was supported in part by the National Science Foundation under Grant No. NSF PHY-1748958.\\    
{\it Software:} NumPy \citep{NumPy}, Matplotlib \citep{Hunter:2007}.
\vspace{-0.5cm}


\section*{Data availability}
The data underlying this article will be shared on reasonable request to the corresponding author.

\bibliographystyle{mnras}
\bibliography{amr} 

\begin{thebibliography}{}
\makeatletter
\relax
\def\mn@urlcharsother{\let\do\@makeother \do\$\do\&\do\#\do\^\do\_\do\%\do\~}
\def\mn@doi{\begingroup\mn@urlcharsother \@ifnextchar [ {\mn@doi@}
  {\mn@doi@[]}}
\def\mn@doi@[#1]#2{\def\@tempa{#1}\ifx\@tempa\@empty \href
  {http://dx.doi.org/#2} {doi:#2}\else \href {http://dx.doi.org/#2} {#1}\fi
  \endgroup}
\def\mn@eprint#1#2{\mn@eprint@#1:#2::\@nil}
\def\mn@eprint@arXiv#1{\href {http://arxiv.org/abs/#1} {{\tt arXiv:#1}}}
\def\mn@eprint@dblp#1{\href {http://dblp.uni-trier.de/rec/bibtex/#1.xml}
  {dblp:#1}}
\def\mn@eprint@#1:#2:#3:#4\@nil{\def\@tempa {#1}\def\@tempb {#2}\def\@tempc
  {#3}\ifx \@tempc \@empty \let \@tempc \@tempb \let \@tempb \@tempa \fi \ifx
  \@tempb \@empty \def\@tempb {arXiv}\fi \@ifundefined
  {mn@eprint@\@tempb}{\@tempb:\@tempc}{\expandafter \expandafter \csname
  mn@eprint@\@tempb\endcsname \expandafter{\@tempc}}}

\bibitem[\protect\citeauthoryear{{Adamo} \& {Bastian}}{{Adamo} \&
  {Bastian}}{2018}]{Adamo_Bastian_18}
{Adamo} A.,  {Bastian} N.,  2018, {The Lifecycle of Clusters in Galaxies}.
p.~91, \mn@doi{10.1007/978-3-319-22801-3_4}

\bibitem[\protect\citeauthoryear{{Adamo} et~al.,}{{Adamo}
  et~al.}{2020}]{Adamo2020}
{Adamo} A.,  et~al., 2020, \mn@doi [\ssr] {10.1007/s11214-020-00690-x}, \href
  {https://ui.adsabs.harvard.edu/abs/2020SSRv..216...69A} {216, 69}

\bibitem[\protect\citeauthoryear{{Bastian}}{{Bastian}}{2008}]{Bastian_08}
{Bastian} N.,  2008, \mn@doi [\mnras] {10.1111/j.1365-2966.2008.13775.x}, \href
  {https://ui.adsabs.harvard.edu/abs/2008MNRAS.390..759B} {390, 759}

\bibitem[\protect\citeauthoryear{{Bastian} \& {Lardo}}{{Bastian} \&
  {Lardo}}{2018}]{bastian18}
{Bastian} N.,  {Lardo} C.,  2018, \mn@doi [\araa]
  {10.1146/annurev-astro-081817-051839}, \href
  {https://ui.adsabs.harvard.edu/abs/2018ARA&A..56...83B} {56, 83}

\bibitem[\protect\citeauthoryear{{Bastian}, {Pfeffer}, {Kruijssen}, {Crain},
  {Trujillo-Gomez}  \& {Reina-Campos}}{{Bastian}
  et~al.}{2020}]{Bastian_et_al_20}
{Bastian} N.,  {Pfeffer} J.,  {Kruijssen} J.~M.~D.,  {Crain} R.~A.,
  {Trujillo-Gomez} S.,   {Reina-Campos} M.,  2020, \mnras~in~press, \href
  {https://ui.adsabs.harvard.edu/abs/2020arXiv200505991B} {p. arXiv:2005.05991}

\bibitem[\protect\citeauthoryear{{Baumgardt}, {Parmentier}, {Anders}  \&
  {Grebel}}{{Baumgardt} et~al.}{2013}]{baumgardt13}
{Baumgardt} H.,  {Parmentier} G.,  {Anders} P.,   {Grebel} E.~K.,  2013,
  \mn@doi [\mnras] {10.1093/mnras/sts667}, \href
  {https://ui.adsabs.harvard.edu/abs/2013MNRAS.430..676B} {430, 676}

\bibitem[\protect\citeauthoryear{{Beasley}, {San Roman}, {Gallart},
  {Sarajedini}  \& {Aparicio}}{{Beasley} et~al.}{2015}]{beasley15}
{Beasley} M.~A.,  {San Roman} I.,  {Gallart} C.,  {Sarajedini} A.,   {Aparicio}
  A.,  2015, \mn@doi [\mnras] {10.1093/mnras/stv943}, \href
  {https://ui.adsabs.harvard.edu/abs/2015MNRAS.451.3400B} {451, 3400}

\bibitem[\protect\citeauthoryear{{Bekki} \& {Tsujimoto}}{{Bekki} \&
  {Tsujimoto}}{2012}]{bekki12}
{Bekki} K.,  {Tsujimoto} T.,  2012, \mn@doi [\apj]
  {10.1088/0004-637X/761/2/180}, \href
  {https://ui.adsabs.harvard.edu/abs/2012ApJ...761..180B} {761, 180}

\bibitem[\protect\citeauthoryear{{Bellstedt} et~al.,}{{Bellstedt}
  et~al.}{2020}]{Bellstedt20}
{Bellstedt} S.,  et~al., 2020, arXiv e-prints, \href
  {https://ui.adsabs.harvard.edu/abs/2020arXiv200511917B} {p. arXiv:2005.11917}

\bibitem[\protect\citeauthoryear{{Choksi}, {Gnedin}  \& {Li}}{{Choksi}
  et~al.}{2018a}]{Choski2018}
{Choksi} N.,  {Gnedin} O.~Y.,   {Li} H.,  2018a, \mn@doi [\mnras]
  {10.1093/mnras/sty1952}, \href
  {https://ui.adsabs.harvard.edu/abs/2018MNRAS.480.2343C} {480, 2343}

\bibitem[\protect\citeauthoryear{{Choksi}, {Gnedin}  \& {Li}}{{Choksi}
  et~al.}{2018b}]{Choksi2018}
{Choksi} N.,  {Gnedin} O.~Y.,   {Li} H.,  2018b, \mn@doi [\mnras]
  {10.1093/mnras/sty1952}, \href
  {https://ui.adsabs.harvard.edu/abs/2018MNRAS.480.2343C} {480, 2343}

\bibitem[\protect\citeauthoryear{{Colucci}, {Bernstein}, {Cameron}  \&
  {McWilliam}}{{Colucci} et~al.}{2011}]{colucci11}
{Colucci} J.~E.,  {Bernstein} R.~A.,  {Cameron} S.~A.,   {McWilliam} A.,  2011,
  \mn@doi [\apj] {10.1088/0004-637X/735/1/55}, \href
  {https://ui.adsabs.harvard.edu/abs/2011ApJ...735...55C} {735, 55}

\bibitem[\protect\citeauthoryear{{Crain} et~al.,}{{Crain}
  et~al.}{2015}]{Crain_et_al_15}
{Crain} R.~A.,  et~al., 2015, \mn@doi [\mnras] {10.1093/mnras/stv725}, \href
  {http://adsabs.harvard.edu/abs/2015MNRAS.450.1937C} {450, 1937}

\bibitem[\protect\citeauthoryear{{Creasey}, {Sales}, {Peng}  \&
  {Sameie}}{{Creasey} et~al.}{2019}]{creasey19}
{Creasey} P.,  {Sales} L.~V.,  {Peng} E.~W.,   {Sameie} O.,  2019, \mn@doi
  [\mnras] {10.1093/mnras/sty2701}, \href
  {https://ui.adsabs.harvard.edu/abs/2019MNRAS.482..219C} {482, 219}

\bibitem[\protect\citeauthoryear{{Da Costa}}{{Da Costa}}{1991}]{da_cost91}
{Da Costa} G.~S.,  1991, in {Haynes} R.,  {Milne} D.,  eds,  IAU Symposium Vol.
  148, The Magellanic Clouds. p.~183

\bibitem[\protect\citeauthoryear{{Da Costa} \& {Hatzidimitriou}}{{Da Costa} \&
  {Hatzidimitriou}}{1998}]{dacosta98}
{Da Costa} G.~S.,  {Hatzidimitriou} D.,  1998, \mn@doi [\aj] {10.1086/300340},
  \href {https://ui.adsabs.harvard.edu/abs/1998AJ....115.1934D} {115, 1934}

\bibitem[\protect\citeauthoryear{{Dalessandro}, {Lapenna}, {Mucciarelli},
  {Origlia}, {Ferraro}  \& {Lanzoni}}{{Dalessandro}
  et~al.}{2016}]{dalessandro16}
{Dalessandro} E.,  {Lapenna} E.,  {Mucciarelli} A.,  {Origlia} L.,  {Ferraro}
  F.~R.,   {Lanzoni} B.,  2016, \mn@doi [\apj] {10.3847/0004-637X/829/2/77},
  \href {https://ui.adsabs.harvard.edu/abs/2016ApJ...829...77D} {829, 77}

\bibitem[\protect\citeauthoryear{{Dias}, {Coelho}, {Barbuy}, {Kerber}  \&
  {Idiart}}{{Dias} et~al.}{2010}]{dias10}
{Dias} B.,  {Coelho} P.,  {Barbuy} B.,  {Kerber} L.,   {Idiart} T.,  2010,
  \mn@doi [\aap] {10.1051/0004-6361/200912894}, \href
  {https://ui.adsabs.harvard.edu/abs/2010A&A...520A..85D} {520, A85}

\bibitem[\protect\citeauthoryear{{Elmegreen} \& {Efremov}}{{Elmegreen} \&
  {Efremov}}{1997}]{elmegreen97}
{Elmegreen} B.~G.,  {Efremov} Y.~N.,  1997, \mn@doi [\apj] {10.1086/303966},
  \href {https://ui.adsabs.harvard.edu/abs/1997ApJ...480..235E} {480, 235}

\bibitem[\protect\citeauthoryear{{Fall} \& {Rees}}{{Fall} \&
  {Rees}}{1985}]{Fall_and_Rees_1985}
{Fall} S.~M.,  {Rees} M.~J.,  1985, \mn@doi [\apj] {10.1086/163585}, \href
  {https://ui.adsabs.harvard.edu/abs/1985ApJ...298...18F} {298, 18}

\bibitem[\protect\citeauthoryear{{Fillingham}, {Cooper}, {Wheeler},
  {Garrison-Kimmel}, {Boylan-Kolchin}  \& {Bullock}}{{Fillingham}
  et~al.}{2015}]{Fillingham_et_al_15}
{Fillingham} S.~P.,  {Cooper} M.~C.,  {Wheeler} C.,  {Garrison-Kimmel} S.,
  {Boylan-Kolchin} M.,   {Bullock} J.~S.,  2015, \mn@doi [\mnras]
  {10.1093/mnras/stv2058}, \href
  {https://ui.adsabs.harvard.edu/abs/2015MNRAS.454.2039F} {454, 2039}

\bibitem[\protect\citeauthoryear{{Forbes}}{{Forbes}}{2020}]{Forbes_20}
{Forbes} D.~A.,  2020, \mn@doi [\mnras] {10.1093/mnras/staa245}, \href
  {https://ui.adsabs.harvard.edu/abs/2020MNRAS.493..847F} {493, 847}

\bibitem[\protect\citeauthoryear{{Forbes} \& {Bridges}}{{Forbes} \&
  {Bridges}}{2010}]{forbes_bridges10}
{Forbes} D.~A.,  {Bridges} T.,  2010, \mn@doi [\mnras]
  {10.1111/j.1365-2966.2010.16373.x}, \href
  {https://ui.adsabs.harvard.edu/abs/2010MNRAS.404.1203F} {404, 1203}

\bibitem[\protect\citeauthoryear{{Forbes} et~al.,}{{Forbes}
  et~al.}{2018}]{forbes_review_18}
{Forbes} D.~A.,  et~al., 2018, \mn@doi [Proceedings of the Royal Society of
  London Series A] {10.1098/rspa.2017.0616}, \href
  {https://ui.adsabs.harvard.edu/abs/2018RSPSA.47470616F} {474, 20170616}

\bibitem[\protect\citeauthoryear{{Gallazzi}, {Charlot}, {Brinchmann}, {White}
  \& {Tremonti}}{{Gallazzi} et~al.}{2005}]{Gallazzi_et_al_05}
{Gallazzi} A.,  {Charlot} S.,  {Brinchmann} J.,  {White} S. D.~M.,   {Tremonti}
  C.~A.,  2005, \mn@doi [\mnras] {10.1111/j.1365-2966.2005.09321.x}, \href
  {https://ui.adsabs.harvard.edu/abs/2005MNRAS.362...41G} {362, 41}

\bibitem[\protect\citeauthoryear{{Garrison-Kimmel} et~al.,}{{Garrison-Kimmel}
  et~al.}{2019}]{Garrison-Kimmel2019}
{Garrison-Kimmel} S.,  et~al., 2019, \mn@doi [\mnras] {10.1093/mnras/stz2507},
  \href {https://ui.adsabs.harvard.edu/abs/2019MNRAS.489.4574G} {489, 4574}

\bibitem[\protect\citeauthoryear{{Geisler}, {Piatti}, {Bica}  \&
  {Clari{\'a}}}{{Geisler} et~al.}{2003}]{geisler03}
{Geisler} D.,  {Piatti} A.~E.,  {Bica} E.,   {Clari{\'a}} J.~J.,  2003, \mn@doi
  [\mnras] {10.1046/j.1365-8711.2003.06408.x}, \href
  {https://ui.adsabs.harvard.edu/abs/2003MNRAS.341..771G} {341, 771}

\bibitem[\protect\citeauthoryear{{Girardi} et~al.,}{{Girardi}
  et~al.}{2013}]{girardi13}
{Girardi} L.,  et~al., 2013, \mn@doi [\mnras] {10.1093/mnras/stt433}, \href
  {https://ui.adsabs.harvard.edu/abs/2013MNRAS.431.3501G} {431, 3501}

\bibitem[\protect\citeauthoryear{{Glatt} et~al.,}{{Glatt}
  et~al.}{2008}]{glatt08}
{Glatt} K.,  et~al., 2008, \mn@doi [\aj] {10.1088/0004-6256/136/4/1703}, \href
  {https://ui.adsabs.harvard.edu/abs/2008AJ....136.1703G} {136, 1703}

\bibitem[\protect\citeauthoryear{{Goudfrooij} et~al.,}{{Goudfrooij}
  et~al.}{2014}]{goudfrooij14}
{Goudfrooij} P.,  et~al., 2014, \mn@doi [\apj] {10.1088/0004-637X/797/1/35},
  \href {https://ui.adsabs.harvard.edu/abs/2014ApJ...797...35G} {797, 35}

\bibitem[\protect\citeauthoryear{{Grcevich} \& {Putman}}{{Grcevich} \&
  {Putman}}{2009}]{Grcevich_and_Putman_09}
{Grcevich} J.,  {Putman} M.~E.,  2009, \mn@doi [\apj]
  {10.1088/0004-637X/696/1/385}, \href
  {https://ui.adsabs.harvard.edu/abs/2009ApJ...696..385G} {696, 385}

\bibitem[\protect\citeauthoryear{{Guglielmo}, {Lewis}  \&
  {Bland-Hawthorn}}{{Guglielmo} et~al.}{2014}]{Guglielmo2014}
{Guglielmo} M.,  {Lewis} G.~F.,   {Bland-Hawthorn} J.,  2014, \mn@doi [\mnras]
  {10.1093/mnras/stu1549}, \href
  {https://ui.adsabs.harvard.edu/abs/2014MNRAS.444.1759G} {444, 1759}

\bibitem[\protect\citeauthoryear{{Hill}}{{Hill}}{1999}]{hill99}
{Hill} V.,  1999, \aap, \href
  {https://ui.adsabs.harvard.edu/abs/1999A&A...345..430H} {345, 430}

\bibitem[\protect\citeauthoryear{{Hollyhead} et~al.,}{{Hollyhead}
  et~al.}{2019}]{Hollyhead19}
{Hollyhead} K.,  et~al., 2019, \mn@doi [\mnras] {10.1093/mnras/stz317}, \href
  {https://ui.adsabs.harvard.edu/abs/2019MNRAS.484.4718H} {484, 4718}

\bibitem[\protect\citeauthoryear{{Hopkins} et~al.,}{{Hopkins}
  et~al.}{2018}]{Hopkins2018}
{Hopkins} P.~F.,  et~al., 2018, \mn@doi [\mnras] {10.1093/mnras/sty1690}, \href
  {https://ui.adsabs.harvard.edu/abs/2018MNRAS.480..800H} {480, 800}

\bibitem[\protect\citeauthoryear{{Horta} et~al.,}{{Horta}
  et~al.}{2020a}]{Horta2020b}
{Horta} D.,  et~al., 2020a, arXiv e-prints, \href
  {https://ui.adsabs.harvard.edu/abs/2020arXiv200710374H} {p. arXiv:2007.10374}

\bibitem[\protect\citeauthoryear{{Horta} et~al.,}{{Horta}
  et~al.}{2020b}]{Horta2020c}
{Horta} D.,  et~al., 2020b, arXiv e-prints, \href
  {https://ui.adsabs.harvard.edu/abs/2020arXiv200801097H} {p. arXiv:2008.01097}

\bibitem[\protect\citeauthoryear{{Horta} et~al.,}{{Horta}
  et~al.}{2020c}]{Horta2020}
{Horta} D.,  et~al., 2020c, \mn@doi [\mnras] {10.1093/mnras/staa478}, \href
  {https://ui.adsabs.harvard.edu/abs/2020MNRAS.493.3363H} {493, 3363}

\bibitem[\protect\citeauthoryear{{Hughes}, {Pfeffer}, {Martig}, {Bastian},
  {Crain}, {Kruijssen}  \& {Reina-Campos}}{{Hughes}
  et~al.}{2019}]{Hughes_et_al_19}
{Hughes} M.~E.,  {Pfeffer} J.,  {Martig} M.,  {Bastian} N.,  {Crain} R.~A.,
  {Kruijssen} J.~M.~D.,   {Reina-Campos} M.,  2019, \mn@doi [\mnras]
  {10.1093/mnras/sty2889}, \href
  {http://adsabs.harvard.edu/abs/2019MNRAS.482.2795H} {482, 2795}

\bibitem[\protect\citeauthoryear{Hunter}{Hunter}{2007}]{Hunter:2007}
Hunter J.~D.,  2007, \mn@doi [Computing In Science \& Engineering]
  {10.1109/MCSE.2007.55}, 9, 90

\bibitem[\protect\citeauthoryear{{Johnson}, {Ivans}  \& {Stetson}}{{Johnson}
  et~al.}{2006}]{johnson06}
{Johnson} J.~A.,  {Ivans} I.~I.,   {Stetson} P.~B.,  2006, \mn@doi [\apj]
  {10.1086/498882}, \href
  {https://ui.adsabs.harvard.edu/abs/2006ApJ...640..801J} {640, 801}

\bibitem[\protect\citeauthoryear{{Kamann} et~al.,}{{Kamann}
  et~al.}{2018}]{kamann18}
{Kamann} S.,  et~al., 2018, \mn@doi [\mnras] {10.1093/mnras/sty1958}, \href
  {https://ui.adsabs.harvard.edu/abs/2018MNRAS.480.1689K} {480, 1689}

\bibitem[\protect\citeauthoryear{{Keller}, {Kruijssen}, {Pfeffer},
  {Reina-Campos}, {Bastian}, {Trujillo-Gomez}, {Hughes}  \& {Crain}}{{Keller}
  et~al.}{2020}]{Keller2020}
{Keller} B.~W.,  {Kruijssen} J.~M.~D.,  {Pfeffer} J.,  {Reina-Campos} M.,
  {Bastian} N.,  {Trujillo-Gomez} S.,  {Hughes} M.~E.,   {Crain} R.~A.,  2020,
  \mn@doi [\mnras] {10.1093/mnras/staa1439}, \href
  {https://ui.adsabs.harvard.edu/abs/2020MNRAS.495.4248K} {495, 4248}

\bibitem[\protect\citeauthoryear{{Kerber}, {Santiago}  \& {Brocato}}{{Kerber}
  et~al.}{2007}]{kerber07}
{Kerber} L.~O.,  {Santiago} B.~X.,   {Brocato} E.,  2007, \mn@doi [\aap]
  {10.1051/0004-6361:20066128}, \href
  {https://ui.adsabs.harvard.edu/abs/2007A&A...462..139K} {462, 139}

\bibitem[\protect\citeauthoryear{{Kim} et~al.,}{{Kim} et~al.}{2018}]{Kim2018}
{Kim} J.-h.,  et~al., 2018, \mn@doi [\mnras] {10.1093/mnras/stx2994}, \href
  {https://ui.adsabs.harvard.edu/abs/2018MNRAS.474.4232K} {474, 4232}

\bibitem[\protect\citeauthoryear{{Kimm}, {Cen}, {Rosdahl}  \& {Yi}}{{Kimm}
  et~al.}{2016}]{kimm2016}
{Kimm} T.,  {Cen} R.,  {Rosdahl} J.,   {Yi} S.~K.,  2016, \mn@doi [\apj]
  {10.3847/0004-637X/823/1/52}, \href
  {https://ui.adsabs.harvard.edu/abs/2016ApJ...823...52K} {823, 52}

\bibitem[\protect\citeauthoryear{{Kruijssen}}{{Kruijssen}}{2012}]{Kruijssen_12}
{Kruijssen} J.~M.~D.,  2012, \mn@doi [\mnras]
  {10.1111/j.1365-2966.2012.21923.x}, \href
  {http://adsabs.harvard.edu/abs/2012MNRAS.426.3008K} {426, 3008}

\bibitem[\protect\citeauthoryear{{Kruijssen}}{{Kruijssen}}{2015}]{Kruijssen_15}
{Kruijssen} J.~M.~D.,  2015, \mn@doi [\mnras] {10.1093/mnras/stv2026}, \href
  {http://adsabs.harvard.edu/abs/2015MNRAS.454.1658K} {454, 1658}

\bibitem[\protect\citeauthoryear{{Kruijssen}, {Pelupessy}, {Lamers}, {Portegies
  Zwart}  \& {Icke}}{{Kruijssen} et~al.}{2011}]{Kruijssen_et_al_11}
{Kruijssen} J.~M.~D.,  {Pelupessy} F.~I.,  {Lamers} H.~J.~G.~L.~M.,  {Portegies
  Zwart} S.~F.,   {Icke} V.,  2011, \mn@doi [\mnras]
  {10.1111/j.1365-2966.2011.18467.x}, \href
  {http://adsabs.harvard.edu/abs/2011MNRAS.414.1339K} {414, 1339}

\bibitem[\protect\citeauthoryear{{Kruijssen}, {Pfeffer}, {Crain}  \&
  {Bastian}}{{Kruijssen} et~al.}{2019a}]{Kruijssen_et_al_19a}
{Kruijssen} J.~M.~D.,  {Pfeffer} J.~L.,  {Crain} R.~A.,   {Bastian} N.,  2019a,
  \mn@doi [\mnras] {10.1093/mnras/stz968}, \href
  {https://ui.adsabs.harvard.edu/abs/2019MNRAS.486.3134K} {486, 3134}

\bibitem[\protect\citeauthoryear{{Kruijssen}, {Pfeffer}, {Reina-Campos},
  {Crain}  \& {Bastian}}{{Kruijssen} et~al.}{2019b}]{Kruijssen_et_al_19b}
{Kruijssen} J.~M.~D.,  {Pfeffer} J.~L.,  {Reina-Campos} M.,  {Crain} R.~A.,
  {Bastian} N.,  2019b, \mn@doi [\mnras] {10.1093/mnras/sty1609}, \href
  {https://ui.adsabs.harvard.edu/abs/2019MNRAS.486.3180K} {486, 3180}

\bibitem[\protect\citeauthoryear{{Kruijssen} et~al.,}{{Kruijssen}
  et~al.}{2020}]{Kruijssen_et_al_20}
{Kruijssen} J.~M.~D.,  et~al., 2020, \mnras~in~press

\bibitem[\protect\citeauthoryear{{Krumholz}, {McKee}  \& {Bland
  -Hawthorn}}{{Krumholz} et~al.}{2019}]{Krumholz19}
{Krumholz} M.~R.,  {McKee} C.~F.,   {Bland -Hawthorn} J.,  2019, \mn@doi
  [\araa] {10.1146/annurev-astro-091918-104430}, \href
  {https://ui.adsabs.harvard.edu/abs/2019ARA&A..57..227K} {57, 227}

\bibitem[\protect\citeauthoryear{{Lah{\'e}n}, {Naab}, {Johansson}, {Elmegreen},
  {Hu}, {Walch}, {Steinwand el}  \& {Moster}}{{Lah{\'e}n}
  et~al.}{2020}]{lahen20}
{Lah{\'e}n} N.,  {Naab} T.,  {Johansson} P.~H.,  {Elmegreen} B.,  {Hu} C.-Y.,
  {Walch} S.,  {Steinwand el} U.~P.,   {Moster} B.~P.,  2020, \mn@doi [\apj]
  {10.3847/1538-4357/ab7190}, \href
  {https://ui.adsabs.harvard.edu/abs/2020ApJ...891....2L} {891, 2}

\bibitem[\protect\citeauthoryear{{Leaman}, {VandenBerg}  \& {Mendel}}{{Leaman}
  et~al.}{2013}]{leaman13}
{Leaman} R.,  {VandenBerg} D.~A.,   {Mendel} J.~T.,  2013, \mn@doi [\mnras]
  {10.1093/mnras/stt1540}, \href
  {https://ui.adsabs.harvard.edu/abs/2013MNRAS.436..122L} {436, 122}

\bibitem[\protect\citeauthoryear{{Li}, {Gnedin}, {Gnedin}, {Meng}, {Semenov}
  \& {Kravtsov}}{{Li} et~al.}{2017}]{li2017}
{Li} H.,  {Gnedin} O.~Y.,  {Gnedin} N.~Y.,  {Meng} X.,  {Semenov} V.~A.,
  {Kravtsov} A.~V.,  2017, \mn@doi [\apj] {10.3847/1538-4357/834/1/69}, \href
  {https://ui.adsabs.harvard.edu/abs/2017ApJ...834...69L} {834, 69}

\bibitem[\protect\citeauthoryear{{Li}, {Gnedin}  \& {Gnedin}}{{Li}
  et~al.}{2018}]{Li2018}
{Li} H.,  {Gnedin} O.~Y.,   {Gnedin} N.~Y.,  2018, \mn@doi [\apj]
  {10.3847/1538-4357/aac9b8}, \href
  {https://ui.adsabs.harvard.edu/abs/2018ApJ...861..107L} {861, 107}

\bibitem[\protect\citeauthoryear{{Ma} et~al.,}{{Ma} et~al.}{2020}]{Ma2020}
{Ma} X.,  et~al., 2020, \mn@doi [\mnras] {10.1093/mnras/staa527}, \href
  {https://ui.adsabs.harvard.edu/abs/2020MNRAS.493.4315M} {493, 4315}

\bibitem[\protect\citeauthoryear{{Madau}, {Lupi}, {Diemand}, {Burkert}  \&
  {Lin}}{{Madau} et~al.}{2020}]{madau20}
{Madau} P.,  {Lupi} A.,  {Diemand} J.,  {Burkert} A.,   {Lin} D. N.~C.,  2020,
  \mn@doi [\apj] {10.3847/1538-4357/ab66c6}, \href
  {https://ui.adsabs.harvard.edu/abs/2020ApJ...890...18M} {890, 18}

\bibitem[\protect\citeauthoryear{{Mar{\'\i}n-Franch}
  et~al.,}{{Mar{\'\i}n-Franch} et~al.}{2009}]{Marin-Franch_et_al_09}
{Mar{\'\i}n-Franch} A.,  et~al., 2009, \mn@doi [\apj]
  {10.1088/0004-637X/694/2/1498}, \href
  {https://ui.adsabs.harvard.edu/abs/2009ApJ...694.1498M} {694, 1498}

\bibitem[\protect\citeauthoryear{{Martocchia} et~al.,}{{Martocchia}
  et~al.}{2019}]{Martocchia19}
{Martocchia} S.,  et~al., 2019, \mn@doi [\mnras] {10.1093/mnras/stz1596}, \href
  {https://ui.adsabs.harvard.edu/abs/2019MNRAS.487.5324M} {487, 5324}

\bibitem[\protect\citeauthoryear{{Massari}, {Koppelman}  \& {Helmi}}{{Massari}
  et~al.}{2019}]{Massari_et_al_19}
{Massari} D.,  {Koppelman} H.~H.,   {Helmi} A.,  2019, \mn@doi [\aap]
  {10.1051/0004-6361/201936135}, \href
  {https://ui.adsabs.harvard.edu/abs/2019A&A...630L...4M} {630, L4}

\bibitem[\protect\citeauthoryear{{McConnachie}}{{McConnachie}}{2012}]{mcconnachie12}
{McConnachie} A.~W.,  2012, \mn@doi [\aj] {10.1088/0004-6256/144/1/4}, \href
  {https://ui.adsabs.harvard.edu/abs/2012AJ....144....4M} {144, 4}

\bibitem[\protect\citeauthoryear{{Mistani} et~al.,}{{Mistani}
  et~al.}{2016}]{Mistani_et_al_16}
{Mistani} P.~A.,  et~al., 2016, \mn@doi [\mnras] {10.1093/mnras/stv2435}, \href
  {https://ui.adsabs.harvard.edu/abs/2016MNRAS.455.2323M} {455, 2323}

\bibitem[\protect\citeauthoryear{{Mucciarelli}, {Carretta}, {Origlia}  \&
  {Ferraro}}{{Mucciarelli} et~al.}{2008}]{mucciarelli08}
{Mucciarelli} A.,  {Carretta} E.,  {Origlia} L.,   {Ferraro} F.~R.,  2008,
  \mn@doi [\aj] {10.1088/0004-6256/136/1/375}, \href
  {https://ui.adsabs.harvard.edu/abs/2008AJ....136..375M} {136, 375}

\bibitem[\protect\citeauthoryear{{Mucciarelli} et~al.,}{{Mucciarelli}
  et~al.}{2011}]{mucciarelli11}
{Mucciarelli} A.,  et~al., 2011, \mn@doi [\mnras]
  {10.1111/j.1365-2966.2010.18167.x}, \href
  {https://ui.adsabs.harvard.edu/abs/2011MNRAS.413..837M} {413, 837}

\bibitem[\protect\citeauthoryear{{Mucciarelli}, {Origlia}, {Ferraro},
  {Bellazzini}  \& {Lanzoni}}{{Mucciarelli} et~al.}{2012}]{mucciarelli12}
{Mucciarelli} A.,  {Origlia} L.,  {Ferraro} F.~R.,  {Bellazzini} M.,
  {Lanzoni} B.,  2012, \mn@doi [\apjl] {10.1088/2041-8205/746/2/L19}, \href
  {https://ui.adsabs.harvard.edu/abs/2012ApJ...746L..19M} {746, L19}

\bibitem[\protect\citeauthoryear{{Mucciarelli}, {Dalessandro}, {Ferraro},
  {Origlia}  \& {Lanzoni}}{{Mucciarelli} et~al.}{2014}]{mucciarelli14}
{Mucciarelli} A.,  {Dalessandro} E.,  {Ferraro} F.~R.,  {Origlia} L.,
  {Lanzoni} B.,  2014, \mn@doi [\apjl] {10.1088/2041-8205/793/1/L6}, \href
  {https://ui.adsabs.harvard.edu/abs/2014ApJ...793L...6M} {793, L6}

\bibitem[\protect\citeauthoryear{{Muratov} \& {Gnedin}}{{Muratov} \&
  {Gnedin}}{2010}]{muratov10}
{Muratov} A.~L.,  {Gnedin} O.~Y.,  2010, \mn@doi [\apj]
  {10.1088/0004-637X/718/2/1266}, \href
  {https://ui.adsabs.harvard.edu/abs/2010ApJ...718.1266M} {718, 1266}

\bibitem[\protect\citeauthoryear{Oliphant}{Oliphant}{06  }]{NumPy}
Oliphant T.,  2006--, {NumPy}: A guide to {NumPy}, USA: Trelgol Publishing,
  \url {http://www.numpy.org/}

\bibitem[\protect\citeauthoryear{{Olsen}, {Hodge}, {Mateo}, {Olszewski},
  {Schommer}, {Suntzeff}  \& {Walker}}{{Olsen} et~al.}{1998}]{olsen98}
{Olsen} K.~A.~G.,  {Hodge} P.~W.,  {Mateo} M.,  {Olszewski} E.~W.,  {Schommer}
  R.~A.,  {Suntzeff} N.~B.,   {Walker} A.~R.,  1998, \mn@doi [\mnras]
  {10.1046/j.1365-8711.1998.01860.x}, \href
  {https://ui.adsabs.harvard.edu/abs/1998MNRAS.300..665O} {300, 665}

\bibitem[\protect\citeauthoryear{{Palma}, {Clari{\'a}}, {Geisler}, {Piatti}  \&
  {Ahumada}}{{Palma} et~al.}{2013}]{palma13}
{Palma} T.,  {Clari{\'a}} J.~J.,  {Geisler} D.,  {Piatti} A.~E.,   {Ahumada}
  A.~V.,  2013, \mn@doi [\aap] {10.1051/0004-6361/201220786}, \href
  {https://ui.adsabs.harvard.edu/abs/2013A&A...555A.131P} {555, A131}

\bibitem[\protect\citeauthoryear{{Parisi} et~al.,}{{Parisi}
  et~al.}{2014}]{parisi14}
{Parisi} M.~C.,  et~al., 2014, \mn@doi [\aj] {10.1088/0004-6256/147/4/71},
  \href {https://ui.adsabs.harvard.edu/abs/2014AJ....147...71P} {147, 71}

\bibitem[\protect\citeauthoryear{{Patel}, {Besla}  \& {Sohn}}{{Patel}
  et~al.}{2017}]{Patel_et_al_17}
{Patel} E.,  {Besla} G.,   {Sohn} S.~T.,  2017, \mn@doi [\mnras]
  {10.1093/mnras/stw2616}, \href
  {https://ui.adsabs.harvard.edu/abs/2017MNRAS.464.3825P} {464, 3825}

\bibitem[\protect\citeauthoryear{{Peebles} \& {Dicke}}{{Peebles} \&
  {Dicke}}{1968}]{peebles68}
{Peebles} P.~J.~E.,  {Dicke} R.~H.,  1968, \mn@doi [\apj] {10.1086/149811},
  \href {https://ui.adsabs.harvard.edu/abs/1968ApJ...154..891P} {154, 891}

\bibitem[\protect\citeauthoryear{{Pfeffer}, {Kruijssen}, {Crain}  \&
  {Bastian}}{{Pfeffer} et~al.}{2018}]{Pfeffer_et_al_18}
{Pfeffer} J.,  {Kruijssen} J.~M.~D.,  {Crain} R.~A.,   {Bastian} N.,  2018,
  \mn@doi [\mnras] {10.1093/mnras/stx3124}, \href
  {http://adsabs.harvard.edu/abs/2018MNRAS.475.4309P} {475, 4309}

\bibitem[\protect\citeauthoryear{{Pfeffer}, {Bastian}, {Crain}, {Kruijssen},
  {Hughes}  \& {Reina-Campos}}{{Pfeffer} et~al.}{2019a}]{Pfeffer_et_al_19a}
{Pfeffer} J.,  {Bastian} N.,  {Crain} R.~A.,  {Kruijssen} J.~M.~D.,  {Hughes}
  M.~E.,   {Reina-Campos} M.,  2019a, \mn@doi [\mnras] {10.1093/mnras/stz1592},
  \href {https://ui.adsabs.harvard.edu/abs/2019MNRAS.487.4550P} {487, 4550}

\bibitem[\protect\citeauthoryear{{Pfeffer}, {Bastian}, {Kruijssen},
  {Reina-Campos}, {Crain}  \& {Usher}}{{Pfeffer}
  et~al.}{2019b}]{Pfeffer_et_al_19b}
{Pfeffer} J.,  {Bastian} N.,  {Kruijssen} J.~M.~D.,  {Reina-Campos} M.,
  {Crain} R.~A.,   {Usher} C.,  2019b, \mn@doi [\mnras]
  {10.1093/mnras/stz2721}, \href
  {https://ui.adsabs.harvard.edu/abs/2019MNRAS.490.1714P} {490, 1714}

\bibitem[\protect\citeauthoryear{{Piatti}, {Hwang}, {Cole}, {Angelo}  \&
  {Emptage}}{{Piatti} et~al.}{2018}]{piatti18}
{Piatti} A.~E.,  {Hwang} N.,  {Cole} A.~A.,  {Angelo} M.~S.,   {Emptage} B.,
  2018, \mn@doi [\mnras] {10.1093/mnras/sty2324}, \href
  {https://ui.adsabs.harvard.edu/abs/2018MNRAS.481...49P} {481, 49}

\bibitem[\protect\citeauthoryear{{Reina-Campos} \& {Kruijssen}}{{Reina-Campos}
  \& {Kruijssen}}{2017}]{Reina-Campos_and_Kruijssen_17}
{Reina-Campos} M.,  {Kruijssen} J.~M.~D.,  2017, \mn@doi [\mnras]
  {10.1093/mnras/stx790}, \href
  {http://adsabs.harvard.edu/abs/2017MNRAS.469.1282R} {469, 1282}

\bibitem[\protect\citeauthoryear{{Reina-Campos}, {Kruijssen}, {Pfeffer},
  {Bastian}  \& {Crain}}{{Reina-Campos} et~al.}{2019}]{Reina-Campos_et_al_19}
{Reina-Campos} M.,  {Kruijssen} J.~M.~D.,  {Pfeffer} J.~L.,  {Bastian} N.,
  {Crain} R.~A.,  2019, \mn@doi [\mnras] {10.1093/mnras/stz1236}, \href
  {https://ui.adsabs.harvard.edu/abs/2019MNRAS.486.5838R} {486, 5838}

\bibitem[\protect\citeauthoryear{{Renzini}}{{Renzini}}{2017}]{renzini17}
{Renzini} A.,  2017, \mn@doi [\mnras] {10.1093/mnrasl/slx057}, \href
  {https://ui.adsabs.harvard.edu/abs/2017MNRAS.469L..63R} {469, L63}

\bibitem[\protect\citeauthoryear{{Rocha}, {Peter}  \& {Bullock}}{{Rocha}
  et~al.}{2012}]{Rocha_et_al_12}
{Rocha} M.,  {Peter} A. H.~G.,   {Bullock} J.,  2012, \mn@doi [\mnras]
  {10.1111/j.1365-2966.2012.21432.x}, \href
  {https://ui.adsabs.harvard.edu/abs/2012MNRAS.425..231R} {425, 231}

\bibitem[\protect\citeauthoryear{{Sabbi} et~al.,}{{Sabbi}
  et~al.}{2007}]{sabbi07}
{Sabbi} E.,  et~al., 2007, \mn@doi [\aj] {10.1086/509257}, \href
  {https://ui.adsabs.harvard.edu/abs/2007AJ....133...44S} {133, 44}

\bibitem[\protect\citeauthoryear{{Saracino} et~al.,}{{Saracino}
  et~al.}{2019}]{saracino19}
{Saracino} S.,  et~al., 2019, \mn@doi [\mnras] {10.1093/mnrasl/slz135}, \href
  {https://ui.adsabs.harvard.edu/abs/2019MNRAS.489L..97S} {489, L97}

\bibitem[\protect\citeauthoryear{{Schaye} et~al.,}{{Schaye}
  et~al.}{2015}]{Schaye_et_al_15}
{Schaye} J.,  et~al., 2015, \mn@doi [\mnras] {10.1093/mnras/stu2058}, \href
  {http://adsabs.harvard.edu/abs/2015MNRAS.446..521S} {446, 521}

\bibitem[\protect\citeauthoryear{{Schechter}}{{Schechter}}{1976}]{Schechter_76}
{Schechter} P.,  1976, \mn@doi [\apj] {10.1086/154079}, \href
  {http://adsabs.harvard.edu/abs/1976ApJ...203..297S} {203, 297}

\bibitem[\protect\citeauthoryear{{Simpson}, {Grand}, {G{\'o}mez}, {Marinacci},
  {Pakmor}, {Springel}, {Campbell}  \& {Frenk}}{{Simpson}
  et~al.}{2018}]{Simpson_et_al_18}
{Simpson} C.~M.,  {Grand} R. J.~J.,  {G{\'o}mez} F.~A.,  {Marinacci} F.,
  {Pakmor} R.,  {Springel} V.,  {Campbell} D. J.~R.,   {Frenk} C.~S.,  2018,
  \mn@doi [\mnras] {10.1093/mnras/sty774}, \href
  {https://ui.adsabs.harvard.edu/abs/2018MNRAS.478..548S} {478, 548}

\bibitem[\protect\citeauthoryear{{Spekkens}, {Urbancic}, {Mason}, {Willman}  \&
  {Aguirre}}{{Spekkens} et~al.}{2014}]{Spekkens_et_al_14}
{Spekkens} K.,  {Urbancic} N.,  {Mason} B.~S.,  {Willman} B.,   {Aguirre}
  J.~E.,  2014, \mn@doi [\apjl] {10.1088/2041-8205/795/1/L5}, \href
  {https://ui.adsabs.harvard.edu/abs/2014ApJ...795L...5S} {795, L5}

\bibitem[\protect\citeauthoryear{{Trenti}, {Padoan}  \& {Jimenez}}{{Trenti}
  et~al.}{2015}]{trenti15}
{Trenti} M.,  {Padoan} P.,   {Jimenez} R.,  2015, \mn@doi [\apjl]
  {10.1088/2041-8205/808/2/L35}, \href
  {https://ui.adsabs.harvard.edu/abs/2015ApJ...808L..35T} {808, L35}

\bibitem[\protect\citeauthoryear{{Usher}, {Brodie}, {Forbes}, {Romanowsky},
  {Strader}, {Pfeffer}  \& {Bastian}}{{Usher} et~al.}{2019}]{Usher_et_al_19}
{Usher} C.,  {Brodie} J.~P.,  {Forbes} D.~A.,  {Romanowsky} A.~J.,  {Strader}
  J.,  {Pfeffer} J.,   {Bastian} N.,  2019, \mn@doi [\mnras]
  {10.1093/mnras/stz2596}, \href
  {https://ui.adsabs.harvard.edu/abs/2019MNRAS.490..491U} {490, 491}

\bibitem[\protect\citeauthoryear{{Wagner-Kaiser} et~al.,}{{Wagner-Kaiser}
  et~al.}{2017}]{wagner-kaiser17}
{Wagner-Kaiser} R.,  et~al., 2017, \mn@doi [\mnras] {10.1093/mnras/stx1702},
  \href {https://ui.adsabs.harvard.edu/abs/2017MNRAS.471.3347W} {471, 3347}

\bibitem[\protect\citeauthoryear{{Weisz}, {Dolphin}, {Skillman}, {Holtzman},
  {Dalcanton}, {Cole}  \& {Neary}}{{Weisz} et~al.}{2013}]{weisz_et_al13}
{Weisz} D.~R.,  {Dolphin} A.~E.,  {Skillman} E.~D.,  {Holtzman} J.,
  {Dalcanton} J.~J.,  {Cole} A.~A.,   {Neary} K.,  2013, \mn@doi [\mnras]
  {10.1093/mnras/stt165}, \href
  {https://ui.adsabs.harvard.edu/abs/2013MNRAS.431..364W} {431, 364}

\bibitem[\protect\citeauthoryear{{Wheeler}, {Phillips}, {Cooper},
  {Boylan-Kolchin}  \& {Bullock}}{{Wheeler} et~al.}{2014}]{Wheeler_et_al_14}
{Wheeler} C.,  {Phillips} J.~I.,  {Cooper} M.~C.,  {Boylan-Kolchin} M.,
  {Bullock} J.~S.,  2014, \mn@doi [\mnras] {10.1093/mnras/stu965}, \href
  {https://ui.adsabs.harvard.edu/abs/2014MNRAS.442.1396W} {442, 1396}

\bibitem[\protect\citeauthoryear{{Wiersma}, {Schaye}  \& {Smith}}{{Wiersma}
  et~al.}{2009a}]{Wiersma_Schaye_and_Smith_09}
{Wiersma} R.~P.~C.,  {Schaye} J.,   {Smith} B.~D.,  2009a, \mn@doi [\mnras]
  {10.1111/j.1365-2966.2008.14191.x}, \href
  {http://adsabs.harvard.edu/abs/2009MNRAS.393...99W} {393, 99}

\bibitem[\protect\citeauthoryear{{Wiersma}, {Schaye}, {Theuns}, {Dalla Vecchia}
   \& {Tornatore}}{{Wiersma} et~al.}{2009b}]{Wiersma_et_al_09}
{Wiersma} R.~P.~C.,  {Schaye} J.,  {Theuns} T.,  {Dalla Vecchia} C.,
  {Tornatore} L.,  2009b, \mn@doi [\mnras] {10.1111/j.1365-2966.2009.15331.x},
  \href {http://adsabs.harvard.edu/abs/2009MNRAS.399..574W} {399, 574}

\makeatother
\end{thebibliography}

\appendix
\section{Observational Sample}
\label{sec:obs_sample}
There have been a number of studies on the age-metallicity relation (AMR) of the stellar cluster and field population of the Magellanic Clouds \citep[e.g.,][]{parisi14}.  To our knowledge, there has not been a systematic study of the ages and metallicities for a large sample of clusters, particularly at the higher mass ($>10^4~\msun$) end.  We have constructed a sample from the literature from a variety of sources, and as such, the sample is heterogeneous.  Ideally, we would like all clusters to have ages determined through multi-band HST CMD analyses, metallicities from medium/high-resolution spectroscopy of large numbers of stars, and masses determined through dynamical and/or profile measurements.  All of these criteria together are rarely met in the literature, hence accommodations must be made.  We selected clusters with a reasonable chance of having present day masses in excess of $10^{4}~\msun$.

In Table~\ref{tab:obs_sample} we list the sample of clusters, along with the adopted parameters and associated references.  We note that this sample is incomplete (even at the relatively high masses, $>10^4~\msun$ used in the present work).  Hence, it is not possible to use this sample for quantitative analysis of the relative age, mass or metallicity distributions of the clusters.  However, it should be representative of the trajectory in AMR space for massive clusters in both the Large and Small Magellanic Clouds.

A systematic study of a representative number of stars with medium/high resolution spectroscopy to determine their abundances (iron and $\alpha$ content, as well as multiple population properties), along with high precision (likely HST) photometry (as well as total stellar mass estimates) to systematically determine the cluster ages would be highly beneficial to the community.

\begin{table*} 

  \begin{tabular}{ |p{1cm}|p{1.5cm}|p{1cm}|p{1cm}|p{3cm}|p{4cm}}
   Galaxy & Cluster &  Age [Gyr] & [Fe/H] & Confidence indication & Reference \\
\hline

LMC & NGC~1466 & 12.2 & -1.7 & 1 &  \citealt[][]{wagner-kaiser17} (WK17)\\
LMC & NGC~1651  & 2.0  & -0.7  &0 &  \citealt[][]{kerber07} (K07)   \\
LMC & NGC~1718  & 2.0  & -0.4  & 0 & K07    \\
LMC & NGC~1777  & 1.1  & -0.6  & 0 & K07    \\
LMC & NGC~1783  & 1.7  & -0.35  & 1 &  \citealt[][]{mucciarelli08}     \\
LMC & NGC~1806  & 1.5  & -0.6  &  1 &  \citealt[][]{mucciarelli14}    \\
LMC & NGC~1831  & 0.7  & -0.1  & 0 & K07    \\
LMC & NGC~1835 & 13.5$^\dagger$  &  -1.8    & 0&   \citealt[][]{olsen98}    \\
LMC & NGC~1841 & 12.6 & -2.0 & 1 &  WK17\\
LMC & NGC~1856  & 0.3  & -0.4  & 0 & K07    \\
LMC & NGC~1866  & 0.18   & -0.4  & 1 &  \citealt[][]{mucciarelli11}, BSV13     \\
LMC & NGC~1868  & 1.1  & -0.7  & 0 & K07    \\
LMC & NGC~1898  & 13.5$^\dagger$  &  -1.2     &  0&  \citealt[][]{johnson06}     \\
LMC & NGC~1916 & 13.5$^\dagger$  &  -1.5 & 0&  \citealt[][]{colucci11}\\
LMC & NGC~2005  & 13.5$^\dagger$  &  -1.8  & 0 &   \citealt[][]{johnson06} \\
LMC & NGC~2019  & 13.5$^\dagger$  &   -1.4  & 0 & \citealt[][]{johnson06}    \\
LMC & NGC~2121  & 2.9  & -0.4  & 0 & K07    \\
LMC & NGC~2136  & 0.09   & -0.4  & 1 &   \citealt[][]{mucciarelli12}   \\
LMC & NGC~2137  & 0.09  & -0.4  &  1 &  \citealt[][]{mucciarelli12}     \\
LMC & NGC~2155  & 2.5  & -0.35  & 1 &  \citealt[][]{Martocchia19}    \\
LMC & NGC~2162  & 1.2  & -0.4  & 0 & K07    \\
LMC & NGC~2173  & 1.6  & -0.6  & 0 & K07    \\
LMC & NGC~2209  & 1.2  & -0.5  & 0 & K07    \\
LMC & NGC~2210 & 10.4 & -1.45 & 1 &  WK17 \\
LMC & NGC~2213  & 1.7  & -0.7  & 0 & K07    \\
LMC & NGC~2249  & 1.0  & -0.4  &0 & K07    \\
LMC & NGC~2257 & 11.5 & -1.7 & 1 &  WK17\\
LMC & ESO~121  & 8.5  & -0.9  & 1 &  \citealt[][]{palma13}     \\
LMC & Hodge~6  & 2.0  & -0.35  &1 &  \citealt[][]{Hollyhead19}    \\
LMC & Hodge~11 & 12.7 & -1.8 & 1 &  WK17\\
LMC & Reticulum & 11.9 & -1.47 & 1 &  WK17 \\
LMC & SL~506  & 2.2  & -0.4  & 0 & K07    \\
LMC & SL~663  & 3.1  & -0.7  & 0 &K07    \\

\hline

SMC & NGC~121  & 10.5  & -1.3  & 1 & \citealt[][]{dalessandro16}    \\
SMC & NGC~330  & 0.03  & -0.8  & 1 & \citealt[][]{hill99}    \\
SMC & NGC~339  & 6.2  & -1.1  & 1 & G08     \\
SMC & NGC~361  & 8.1  & -1.0  & 0 & D10    \\
SMC & NGC~411  & 1.7  & -0.8  &1 & \citealt[][]{girardi13}    \\
SMC & NGC~416  & 6.2  & -1.0  & 1 & G08     \\
SMC & NGC~419  & 1.5  & -0.7   & 1 & G08     \\
SMC & BS-10  & 4.5  & -1.0  & 1 & \citealt[][]{sabbi07}   \\
SMC & BS-90  & 4.3  & -1.0  &  0 &  D10  \\
SMC & Kron~3  & 6.8  & -1.1  & 1 & G08    \\
SMC & Kron~23  & 2.1  & -1.2  &  0 &  G08, \citealt[][]{dias10} (D10) \\
SMC & Kron~44  & 3.1  & -1.1  &   0 &  D10 \\
SMC & Lindsay~1  & 8  & -1.3  & 1 & \citealt[][]{glatt08} (G08) \\
SMC & Lindsay~11  & 3.5  & -0.8  & 0 & D10    \\
SMC & Lindsay~32  & 4.8  & -1.2  & 0 & D10    \\
SMC & Lindsay~113  & 4.0  & -1.2  &1 &  \citealt[][]{dacosta98} \\
SMC & Lindsay~116  & 2.8  & -1.1  &0 &  D10     \\

    \hline 

    \hline 
  \end{tabular}
\caption{The adopted properties of massive clusters in the LMC and SMC.  Additionally, we give a confidence indication, with 1 being high confidence and 0 being lower confidence on (at least) one of the cluster properties (age, [Fe/H]).  $^\dagger$ indicates clusters without a systematic precision age measurement, which are given an arbitrary ancient age of 13.5~Gyr. }
\label{tab:obs_sample}
\end{table*}

\bsp	
\label{lastpage}
\end{document}